\documentclass[pra,longbibliography,twocolumn,showpacs,superscriptaddress,notitlepage,floatfix]{revtex4-2}

\usepackage{amsmath}
\usepackage{amssymb,bm}
\usepackage{amsthm}
\usepackage{newfloat}
\usepackage{algpseudocode}
\DeclareFloatingEnvironment[fileext=loa,listname=List of Algorithms,name=Algorithm]{algorithm}
\makeatletter
\renewcommand{\fnum@algorithm}{\algorithmname\ \thealgorithm}
\makeatother

\usepackage{qcircuit}
\usepackage{nicematrix}
\newtheorem{innercustomgeneric}{\customgenericname}
\providecommand{\customgenericname}{}
\newcommand{\newcustomtheorem}[2]{%
  \newenvironment{#1}[1]
  {%
   \renewcommand\customgenericname{#2}%
   \renewcommand\theinnercustomgeneric{##1}%
   \innercustomgeneric
  }
  {\endinnercustomgeneric}
}

\newcustomtheorem{customthm}{Theorem}
\newcustomtheorem{definitionBob}{Definition}

\usepackage{color,dsfont} 
\usepackage{graphicx}
\usepackage{subcaption}
\usepackage{ragged2e}
\DeclareCaptionJustification{justified}{\justifying}
\definecolor{linkColor}{rgb}{0.6,0.44,0.67}

\usepackage[colorlinks=true, hyperindex, breaklinks, allcolors=linkColor]{hyperref} 
\usepackage[normalem]{ulem}
\usepackage{zref-clever}
\zcsetup{cap}
\usepackage{mathrsfs}

\usepackage{mathtools}
\usepackage{float}
\usepackage{verbatim}
\usepackage{latexsym}
\usepackage{amsmath}
\usepackage{amssymb}
\usepackage{setspace}
\usepackage{amsfonts}
\usepackage{stmaryrd}
\usepackage{xcolor}
\usepackage{enumitem}
\usepackage{hhline}
\usepackage{physics}

\usepackage{soul}

\usepackage{bbm}

\newcommand{\codepar}[1]{\ensuremath{[\![#1]\!]}}
\newcommand{\ccodepar}[1]{\ensuremath{[#1]}}
\newcommand{\gencodepar}[1]{\ensuremath{(\!(#1)\!)}}

\definecolor{azure}{rgb}{0.0, 0.5, 1.0}
\definecolor{blue-green}{rgb}{0.0, 0.67, 0.57}
\definecolor{neoncarrot}{rgb}{1.0, 0.64, 0.26}

\zcRefTypeSetup{equation}{
  Name-sg-ab = {Eq.\!}, 
  name-sg-ab = {eq.\!}, 
  Name-pl-ab = {Eqs.\!}, 
  name-pl-ab = {eqs.\!}, 
  abbrev, 
}
\zcRefTypeSetup{figure}{
  Name-sg-ab = {Fig.}, 
  name-sg-ab = {fig.}, 
  Name-pl-ab = {Figs.}, 
  name-pl-ab = {figs.}, 
  abbrev, 
}
\usepackage{multirow}
\mathchardef\mhyphen="2D

   \allowdisplaybreaks

\begin{document}

\title{Trade-offs in Gauss's law error correction for lattice gauge theory quantum simulations}

\author{Balint Pato}
\email{balint.pato@duke.edu}
\affiliation{
    Duke Quantum Center, Duke University, Durham, NC 27701, USA
}
\affiliation{
    Department of Electrical and Computer Engineering, Duke University, Durham, NC 27708, USA
}
\author{Natalie Klco}
\email{natalie.klco@duke.edu}
\affiliation{
    Duke Quantum Center, Duke University, Durham, NC 27701, USA
}
\affiliation{
    Department of Physics, Duke University, Durham, NC 27708, USA
}

\begin{abstract}
    Gauss's law-based quantum error correction (GLQEC) offers a promising approach to reducing qubit overhead in lattice gauge theory simulations by leveraging built-in symmetries. For applications of GLQEC to 1+1D lattice quantum electrodynamics (QED), we identify two significant trade-offs. First, we prove via dimension-counting arguments that GLQEC requires periodic electric fields, thereby constraining the design space for lattice QED simulations. Second, we numerically compare GLQEC with a universal quantum error correction (UQEC) code, specifically the $d=3$ bitflip repetition code, and find that while GLQEC can achieve lower logical error rates in single-round error correction, it exhibits faster decoherence to the steady-state mixed ensemble under multiple rounds. The mixing speed penalty is manifest in observables of interest for both memory experiments and Hamiltonian evolution. We identify a mixing speed threshold, $p_{th}=0.277(2)$, above which using GLQEC exhibits even faster decoherence than without error correction. Our results highlight fundamental limitations of symmetry-based error correction schemes and inform corresponding constraints on formulations of lattice gauge theories compatible with error-robust quantum simulation techniques.
\end{abstract}

\maketitle
\tableofcontents

\section{Introduction}

Quantum error correction (QEC) is widely considered to be the most promising path towards scaling up quantum computers \cite{shorFaulttolerantQuantumComputation1996}. A typical \codepar{n,k,d} QEC is application-agnostic; it encodes $k$ logical qubits into $n > k$ physical ones, ensuring that the logical operators are non-local and are at least of weight $d$. Applications then operate on these logical qubits, using the logical operators. Stabilizer codes \cite{gottesman_heisenberg_1998} are the most prominent practical family of QEC codes, which work by introducing new symmetries respected by the physical qubits, enforced by stabilizers. However, for certain applications like quantum simulation of lattice gauge theories (LGT)~\cite{Byrnes:2005qx,Zohar:2012ay,Banerjee:2012pg,Zohar:2012xf,Banerjee:2012xg,Zohar:2014qma}, there are ``built-in'' symmetries. When they are not emergent or encoded in hardware~\cite{Zohar:2013zla,Hauke:2013jga,Marcos:2014lda,Zohar:2015hwa}, these symmetries can be leveraged as partial QEC solutions \cite{rajput_quantum_2023,spagnoli_fault-tolerant_2024,carena_quantum_2024}. Well beyond the presence of energetic hierarchies~\cite{Klco:2021jxl} in quantum field simulations, incorporation of these symmetries into QEC strategies leads to a reduction in the qubit overhead required for fault tolerance.

The symmetry to leverage in LGT simulations is Gauss's law. While this symmetry has most commonly been analytically removed from quantum simulation implementations to avoid gauge-violating errors~\cite{hamerSeriesExpansionsMassive1997,martinezRealtimeDynamicsLattice2016d,nguyenDigitalQuantumSimulation2022b,florioRealTimeNonperturbativeDynamics2023,kokailSelfverifyingVariationalQuantum2019,farrellScalableCircuitsPreparing2024,grieningerQuantumComplexityString2026}, it has also been employed as a source of error detection to guide critical post-selections in quantum simulations~\cite{stryker_oracles_2019,klcoSU2NonAbelianGauge2020}. More recently, Gauss's law has been shown to be useful in an error-correction strategy for Abelian LGTs, as Refs.~\cite{rajput_quantum_2023,spagnoli_fault-tolerant_2024} (RRW) introduced a stabilizer code from Gauss's law for the $\mathbb{Z}_2$ and periodic $\Lambda=1$-truncated $U(1)$ theories. For non-Abelian pure gauge LGTs it has been shown that for a specific parameter regime, QEC based on Gauss's law can be more efficient than gauge fixing \cite{carena_gauge_2022}, and explicit constructions for $SU(2)$ theories have been built \cite{yaoQuantumErrorCorrection2025}. Despite the interest in these solutions, there are no studies available on the performance of these codes in simulation, and, to the authors' best knowledge, no rigorous analysis of the logical channel has been conducted.

In this work, we evaluate the RRW protocol of Gauss's Law QEC (GLQEC) for 1+1D lattice QED, using the Kogut-Susskind Hamiltonian \cite{kogut_hamiltonian_1975,banks_strong-coupling_1976} formulation of the Schwinger model \cite{schwinger_gauge_1962}. Firstly, we noticed that in truncated $U(1)$ LGTs, if the electric field is non-periodic, the RRW GLQEC allows extra, unphysical states in the codespace. Using dimensionality analysis, we prove that any binary stabilizer code based on Gauss's law is only compatible with periodic electric fields and their associated modular definition of physicality.

A common way of simulating a QEC solution is the single-round memory experiment under code capacity noise. We find that, compared to a universal QEC (UQEC) setup of the $d=3$ repetition code, under bitflip noise, GLQEC performs slightly better in many physically relevant regimes, with a lower logical error rate. This advantage behaves non-trivially in increasing system size and as a function of the physical error rate. The UQEC-to-GLQEC logical error rate ratio converges to 1.2(1.0) asymptotically at low(high) physical error rates, transitioning in an intermediate regime that is located at reduced error rates with increasing volume.

Next, going beyond the typical QEC benchmarking, we explored how GLQEC performs in a noisy Hamiltonian simulation compared to UQEC, for both the periodic and non-periodic cases. Surprisingly, despite its superior single-round performance, the logical information in GLQEC exhibits faster decoherence than that in UQEC. We also verified this behavior in a multi-round memory experiment. Another interesting effect is that the GLQEC steady-state ensembles differ from those of UQEC. Together, these two effects lead to significant variation in characterizing the relative proximity of GLQEC and UQEC to noiseless evolutions, depending on observable and simulation regime. To understand the origin of the mixing penalty, we analytically investigate rapidly convergent estimates of mixing rates in smaller systems via spectral gap analysis.

The paper is structured as follows. In \zcref{sec: background} we establish our notation and describe different options for deploying the Schwinger model on a quantum computer, basic notions in quantum error correction, and the RRW scheme. In \zcref{sec: dim of schwinger} we derive that the dimension of the $U(1)$ Schwinger model with non-periodic electric field follows the Lucas numbers and thus show the incompatibility with stabilizer codes. We showcase our numerical results in \zcref{sec: simulations}, which include the single round memory experiment where GLQEC outperforms UQEC, and the mixing penalty with and without Hamiltonian evolution. In \zcref{ssec:mixing speed analysis}, we study the mixing speed phenomenon numerically using estimators of the electric energy and the trace distance from the steady state based on the second largest eigenvalue modulus, and by approximate analytical formulas for the second largest eigenvalue modulus of the GLQEC channel. Finally, we discuss our results and conclude in \zcref{sec: discussion}.

\section{Background} \label{sec: background}

In this section, we review the Schwinger model, an Abelian lattice gauge theory, and its mapping to a quantum computer. We then briefly summarize the fundamental concepts of quantum error correction and the RRW scheme for Gauss's law QEC.

\subsection{The massive Schwinger model on a lattice} \label{sec: schwinger model}

The massless \cite{schwinger_gauge_1962} and massive Schwinger models \cite{kogut_how_1975} are simplified, 1+1-dimensional theories of quantum electrodynamics that enable the simplified study of properties, such as confinement and exact gauge invariance, that arise in more complex theories, for example, quantum chromodynamics. Lattice gauge theories \cite{wilsonConfinementQuarks1974} are discretized versions of continuum field theories that converge to the continuum version in the limit of vanishing lattice spacing and diverging correlation lengths such that physical length scales are resolved with increasing precision. In this paper, we explore the lattice gauge theory of the massive Schwinger model in the Kogut-Susskind Hamiltonian formulation \cite{kogut_hamiltonian_1975,banks_strong-coupling_1976}, which uses staggered fermions. Staggered fermions are represented using an alternating interpretation of fermionic occupation states. For even fermionic sites (starting from $s=0$), the presence of an electron is represented by vacuum states, while on odd sites, occupied fermionic states are interpreted as the presence of a positron. A \textit{physical site} thus consists of a pair of neighboring fermion-antifermion sites, and the strong-coupling vacuum of zero particle number is antiferromagnetic. We denote the number of physical sites with $n$ and the number of fermionic sites with $n_f = 2n$. The Hamiltonian is the sum of the hopping term $H_I$, the electric term $H_E$ and the mass term $H_M$:
\begin{align}
    H   & = H_I + H_E + H_M \label{eq: hamiltonian}                                                                          \\
    H_I & =\ x \sum_{s=0}^{n_f-1} \left(\psi_s^\dagger U_{s,s+1}\psi_{s+1} + \psi_s U_{s,s+1}^{\dagger} \psi_{s+1}^{\dagger}
    \right) \label{eq: interaction term}                                                                                     \\
    H_M & =\sum_{s=0}^{n_f-1}  \mu (-)^{s} \psi^\dagger_s \psi_s \label{eq: electric term}                                   \\
    H_E & =\sum_{s=0}^{n_f-1} E_{s,s+1}^2 \label{eq: mass term} \,,
\end{align}
where $\mu = 2m/(ag^2)$ and $x = 1/(ag)^2$, with $m$ the fermionic mass, $g$ the coupling constant, and $a$ the lattice spacing. The operators $\psi^\dagger_s, \psi_s$ are the fermionic creation and annihilation operators on site $s$, respectively, satisfying the fermionic commutation relations:
\begin{align*}
    \{\psi_r^\dagger, \psi_s^\dagger \} =\{\psi_r, \psi_s\} & = 0              \\
    \{\psi_s, \psi_r^\dagger\}                              & = \delta_{rs}\,.
\end{align*}
The charge operator is defined as
%
\begin{align}
    \rho_s =  \psi^\dagger_s \psi_s - \frac{1+(-)^s }{2}  \label{eq: charge op cont}\,,
\end{align}
following the staggered fermion convention. Thus, the site operators act on the fermionic Hilbert space spanned by two basis states, with the vacuum representing the state of lower charge, meaning the presence of an electron on even sites or vacuum (absence of a positron) on odd sites.

For a link connecting sites $s$ and $s+1 \mod n_f$, the gauge Hilbert space is infinite-dimensional with a basis indexed by unitless integers of the electric flux $\ket{-\infty}$ to $\ket{+\infty}$. The two conjugate operators that act on it are the link variable $U_{s,s+1}$, a unitary operator equal to the exponential of the vector potential in the temporal-gauge gauge field $A_\mu=(0,A_1)$, and $E_{s,s+1}$, the electric flux operator:
\begin{align}
    U_{s,s+1} & = e^{iaA_{s,s+1}} = \sum_{m=-\infty}^{\infty} \bigl(\ketbra{m+1}{m}\bigr)_{s,s+1} \\
    E_{s,s+1} & = \sum_{m=-\infty}^{\infty} m \Bigl(\ketbra{m}{m}\Bigr)_{s,s+1} \,.
\end{align}
These operators commute on different links, while on the same link, they obey the following:
\begin{align}
    [E_{s,s+1}, U_{s,s+1}] = U_{s,s+1}\,.
\end{align}
Gauss's law is a local constraint on each site $s$:
\begin{align}
    (\nabla \cdot E) (s) - \rho_s =0\,.
\end{align}
We will express Gauss's law as an operator for each site, which will define the physical subspace:
\begin{align}
    G_{s}              & = E_{s+1, s} - E_{s,s-1}  - \rho_s \label{eq: gauss law op in u(1)} \\
    \mathcal{H}_{phys} & = \bigcap_{s=0}^{n_f-1} \ker G_s \,.
\end{align}

As our quantum computers are finite-dimensional, we are forced to use gauge field discretization to tame the infinite-dimensional gauge field, alongside the spacetime discretization. When considering the continuum limit, convergence must be systematically correlated along both of these axes, approaching both lattice spacing $a=0$ and infinite dimensions in the link Hilbert space. Next, we will discuss two gauge field discretization approaches, the truncated $U(1)$ theories with or without periodicity in the electric field. As the majority of today's quantum computer architectures rely on qubits (less so on qudits), we will also describe the qubit representation of fermionic operators using the Jordan-Wigner transformation \cite{jordanUeberPaulischeAequivalenzverbot1928} and the two-dimensional, qubit truncation of the link operators.

\subsubsection{Finite-dimensional gauge field theories}

The two main options for gauge field discretization we will discuss in this paper are the truncated $U(1)$ theories with a periodic electric field or a non-periodic electric field. $\mathbb{Z}_d$ theories are a popular choice, but, as their behavior is very similar for our purposes to the truncated $U(1)$ theory with periodic electric field, we only discuss them in \zcref{app: modular}. In this subsection, we define the elements of the electric field basis and the link operators that act on the truncated $U(1)$ theories.

Our main choice of discretization is to keep the gauge group $G=U(1)$, but truncate the electric field basis states to a finite order. The basis states are labeled by integers, similar to the infinite-dimensional case. The truncation in general can be symmetric around zero (e.g., in Ref.~\cite{klco_quantum-classical_2018}) or asymmetric (e.g., in the RRW scheme for qubits \cite{rajput_quantum_2023}). In the asymmetric case, with cutoff $\Lambda \in \mathbb{N}_+$, the basis elements will correspond to the flux values $\ket{m} \in \{\ket{-\Lambda}, \ket{-\Lambda+1}, \ldots \ket{\Lambda-1}\}$, in the symmetric case $\ket{m} \in \{\ket{-\Lambda}, \ket{-\Lambda+1}, \ldots \ket{\Lambda-1}, \ket{\Lambda}\}$. Now the remaining choice is whether to make the field periodic. We denote the periodic, $\Lambda$-truncated $U(1)$ theory with $U(1)_{d}^\circ$, while the non-periodic one with $U(1)_d^-$, where $d=2\Lambda+1$ for symmetric and $d=2\Lambda$ for asymmetric theories. The link operators then agree on $E$, but they differ in the $U$ operator, as $U^{U(1)_d^\circ}$ is unitary while $U^{U(1)_d^-}$ is non-unitary due to its annihilation of the basis state with the highest flux value,
\begin{align}
    E^{U(1)_d^-}_{s,s+1}     & =E^{U(1)_d^\circ}_{s,s+1} = \sum_{m=-\Lambda}^{d-\Lambda-1} m \bigl( \ketbra{m}{m}\bigr)_{s,s+1} \\
    U_{s,s+1}^{U(1)_d^-}     & =\!\!\! \sum_{m=-\Lambda}^{d-\Lambda-2} \bigl(\ketbra{m+1}{m}\bigr)_{s,s+1}                      \\
    U_{s,s+1}^{U(1)_d^\circ} & =  U_{s,s+1}^{U(1)_d^-} +\!  \bigl(\ketbra{-\Lambda}{d-\Lambda-1} \bigr)_{s,s+1}\,.
\end{align}
The non-periodic $\Lambda$-truncated $U(1)$ preserves the commutation relations from the non-truncated $U(1)$ theory, while the periodic theory commutation relations are modified at the boundary:
\begin{align}
    \left[E_{s,s+1}^{U(1)_d^-}, U_{s,s+1}^{U(1)_d^-} \right]\!         & = U_{s,s+1}^{U(1)_d^-}                                                       \\
    \left[E_{s,s+1}^{U(1)_d^\circ}, U_{s,s+1}^{U(1)_d^\circ} \right]\! & = U_{s,s+1}^{U(1)_d^\circ}                                                   \\
                                                                       & \ \ + (1-d) \bigl(\ketbra{-\Lambda}{d-\Lambda-1} \bigr)_{s,s+1} \nonumber\,.
\end{align}
For truncated theories with periodic electric fields, the associated Gauss's law will be \zcref{eq: gauss law op in u(1)} evaluated with $d$-modular arithmetic.

When choosing between discretization strategies, beyond the commutation relations and unitarity of the link variable, other implementation concerns on a quantum computer can also inform the decision.

\subsubsection{Mapping the fermionic and link operators to qubits} \label{ssec: qubitized lgt}

Using qubits to represent the states of the fermionic sites is very natural. We can follow the staggered fermion convention by encoding $\ket{0}$ for $e^-$, $\ket{1}$ for vacuum on even sites and the opposite way for odd sites, where $\ket{0}$ denotes the vacuum and $\ket{1}$ the positron.

As for the site operators, we will use the well-known Jordan-Wigner transformation, which preserves the fermionic algebra while acting on qubits with Pauli operators. The creation and annihilation operators will be:
\begin{align}
    \psi_s^\dagger & = \left(\prod_{t=0}^{s-1} Z_t \right) \frac{1}{2}(X_s - iY_s)    \\
    \psi_s         & = \left(\prod_{t=0}^{s-1} Z_t \right) \frac{1}{2}(X_s + iY_s)\,,
\end{align}
where a Pauli operator $P_s$ acts on the qubit corresponding to site $s$. For the charge operator, to represent \zcref{eq: charge op cont}, we will have:
\begin{align}
    \rho_s & = -\frac{1}{2}(Z_s + (-)^{s} I_s) = \begin{cases}
                                                     -\ketbra{0}{0}, \text{ for } s \text { even} \\
                                                     + \ketbra{1}{1}, \text{ for } s \text { odd}
                                                 \end{cases} \,.
\end{align}
For most of this paper, we use qubits to represent both the fermionic modes and a two-dimensional gauge Hilbert space (we will refer to these as \textit{binary lattice gauge theories}). The gauge group is then $G=U(1)$ with an asymmetric flux cutoff $\Lambda=1$. The electric field basis states are represented by $\ket{0}$ for flux $-1$ and by $\ket{1}$ for flux $0$. The qubit link operators for the various theories are as follows:
\begin{align}    E^{U(1)_2^-}_{s,s+1}=E^{U(1)_2^\circ}_{s,s+1} & = - \bigl(\ketbra{0}{0}\bigr)_{s,s+1} \\
                 U_{s,s+1}^{U(1)_d^-}                          & = \bigl(\ketbra{1}{0}\bigr)_{s,s+1}   \\
                 U_{s,s+1}^{U(1)_d^\circ}                      & = X_{s,s+1}\,.
\end{align}
In this representation, fermion and gauge configurations on the lattice are mapped to $N = 4n = 2n_f$ qubits. We will refer to these qubits as \textbf{application qubits} to distinguish from the quantum error corrected \textit{logical qubit} term. Namely, logical qubits can be used to implement application qubits, but in the application-specific scheme described in \zcref{ssec: glqec}, and when no encoding is used, physical qubits are used directly as application qubits.

\subsection{Quantum Error Correction}

The goal of quantum error correction \cite{shorFaulttolerantQuantumComputation1996} (QEC) is to protect quantum information by encoding its operators into non-local operators, acting on a larger Hilbert space. Under the assumption that \textit{noise} (a collection of unwanted physical processes) acts locally, a QEC scheme can suppress the error rate, given the error rate is below the \textit{accuracy threshold} of the code \cite{aharonov_fault-tolerant_2008, dennis_topological_2002, knill_threshold_1996}. An \gencodepar{n,k,d} QEC qubit code is a $2^k$-dimensional subspace of $k$ logical qubits within a $2^n$-dimensional Hilbert space of $n$ physical qubits with logical operators that have to act on at least $d$ qubits (the minimum distance of the code).

The most prominent family of quantum error correcting codes is the family of stabilizer codes \cite{gottesman_heisenberg_1998}, to which most practical QEC codes belong. Stabilizer codes are denoted with the square-bracket notation, \codepar{n,k,d}, to differentiate from the more general, potentially non-additive (non-stabilizer) QEC codes \cite{rains_nonadditive_1997,yu_nonadditive_2008} with the same meaning for $n,k,d$ as above. The code subspace of stabilizer codes is defined by the shared +1 eigenspace of $n-k$ commuting Pauli operators called the stabilizer generators. The stabilizer generators generate the Abelian \textit{stabilizer group}. In a practical setting, this structure allows for an efficient method to extract error information from the system and recover it. This is in contrast with non-additive (non-stabilizer) codes \cite{yu_nonadditive_2008, rains_nonadditive_1997,cross_codeword_2009}, for which, in general, only exponentially expensive measurement and decoding schemes are known \cite{li_clustered_2010}.

Under standard noise assumptions, most protocols are evaluated using stochastic Pauli noise, quantum channels described by Kraus operators that are elements of the Pauli group. The single and two-qubit depolarizing channel, bitflip, and the phase-flip channels are all Pauli channels.

Active QEC protocols based on stabilizer codes repeatedly measure the eigenvalues of the stabilizer generators and calculate recovery based on the results. Each round of measurement yields a length-$(n-k)$ bitstring, the \textit{syndrome}, $\boldsymbol{s}$. A syndrome bit $s_i=1$ when the stabilizer generator $g_i$ anticommutes with the error $E$ the system experienced. Otherwise, the bit will be zero.

The syndrome is then decoded by a classical algorithm, the decoder, to guess the recovery operator $R$. If the final operator $ER$ is an element of the stabilizer group $S$, the recovery succeeded; otherwise, a logical error has occurred.

In the special case of Calderbank-Shor-Steane \cite{shorFaulttolerantQuantumComputation1996,calderbank_good_1996,steane_simple_1996} (CSS) family, the stabilizer generators are all either a tensor product of $X$ or a tensor product of $Z$ operators. This allows for separate decoding of $X$-type or $Z$-type errors. The stabilizer generators (or checks) for \codepar{n,k,d} CSS codes can be represented with two binary matrices, $H_X$ and $H_Z$, each of dimension $(n-k)\times n$ that must satisfy $H_XH_Z^T= 0$ due to the commutativity requirement. In the symplectic representation, we can represent single-qubit Pauli operators with two bits following $I=(0,0), X=(1,0), Z=(0,1), Y=(1,1)$. A Pauli error $E$ can then be decomposed as $E=E_XE_Z \in \mathcal{P}_n$ as a product of $E_X=\bigotimes_{i=1}^n X_i^{a_i}$ and $E_Z=\bigotimes_{i=1}^n Z_i^{b_i}$, for binary strings $a,b \in \mathbb{F}_2^n$. Then the syndrome $\boldsymbol{s}(E)=(\boldsymbol{s_X}(E_Z)|\boldsymbol{s_Z}(E_X))$ can be split up into $X$-syndrome $\boldsymbol{s_X} = H_XE_Z$ and $Z$-syndrome, $\boldsymbol{s_Z}=H_ZE_X$. Using this formalism, in fact, we can separate $X$ and $Z$ distance of a code, $d_X$ and $d_Z$, and then the minimum distance is the minimum of the two, $d = \min(d_X,d_Z)$. Sometimes, we will use \codepar{n,k,d_X:d_Z} notation to emphasize the difference between the two axes.

\begin{table}[htbp!]
    \centering
    \begin{tabular}{c}
        \texttt{ZZZ ZZZ III} \\
        \texttt{III ZZZ ZZZ} \\
        \texttt{XXI III III} \\
        \texttt{IXX III III} \\
        \texttt{III XXI III} \\
        \texttt{III IXX III} \\
        \texttt{III III XXI} \\
        \texttt{III III IXX}
    \end{tabular}
    \caption{Shor's \codepar{9,1,3} code's stabilizer generators. }
    \label{tab: shor stabilizers}
\end{table}

Finally, we introduce the concept of code concatenation for two codes $C_1$ of parameters \codepar{n_1,1,d_{X,1}\!:\!d_{Z,1}} and $C_2$ with parameters \codepar{n_2,k_2,d_{X,2}\!:\!d_{Z,2}}. The concatenated code is a \codepar{n_1n_2, k_2, d_{X,1}d_{X,2}\!:\!d_{Z,1}d_{Z,2}} code, where each physical qubit of the outer code $C_2$ is replaced with the logical qubit of the $C_1$ code. A well-known example of a concatenated code is the \codepar{9,1,3} Shor's code \cite{shorFaulttolerantQuantumComputation1996}, which protects against single-bitflip and phase-flip errors. It is the concatenation of a \codepar{3,1,3\!:\!1} bitflip code and a \codepar{3,1,1\!:\!3} phase-flip code, which are both just \ccodepar{3,1,3} classical codes, but used for protection against $X$ and $Z$ errors, respectively. In the original formulation, the phase-flip code is the outer code, but in our case, we will use the bitflip code as the outer code. This formulation is sometimes called the $X$-Shor code \cite{pato_logical_2025}, as the weight-2 stabilizer generators are of $X$ type. The bitflip code has stabilizer generators $\langle ZZI, IZZ \rangle$ with logical $X_L=XXX$, while the phase-flip code has stabilizer generators $\langle XXI, IXX \rangle$. The final set of 8 stabilizer generators is in \zcref{tab: shor stabilizers}. This code can correct all single-qubit phase-flip and single-qubit bitflip errors.

\subsection{Gauss's law Quantum Error Correction} \label{ssec: glqec}

Here we review the RRW protocol~\cite{rajput_quantum_2023, spagnoli_fault-tolerant_2024} for using Gauss's law for quantum error correction in the $U(1)_2^\circ$ theories.
\begin{figure}[htpb!]
    \begin{subfigure}{1\textwidth}
        \refstepcounter{subfigure}\label{fig:RRW:a}
        \refstepcounter{subfigure}\label{fig:RRW:b}
        \refstepcounter{subfigure}\label{fig:RRW:c}
    \end{subfigure}%
    \centering
    \includegraphics[width=1\linewidth]{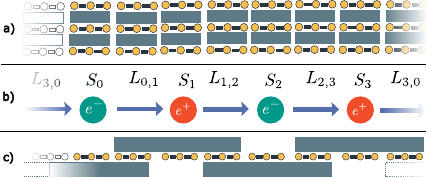}
    \caption{A concatenated universal QEC code and the RRW QEC code. $X$-checks are black and $Z$ checks are dark gray. a) A universal error correction scheme using Shor's \codepar{9,1,3} code for each site and link qubit. b) The corresponding 1+1D lattice of four staggered fermionic sites $S_s$ connected by links $L_{s,s+1}$  with periodic boundary conditions. c) The RRW scheme \cite{rajput_quantum_2023} using a $d=3$ phase-flip code concatenated with the 3-body Gauss's law checks ($Z^{\otimes 9}$ checks).}
    \label{fig:rrw lattice}
\end{figure}
In binary lattice gauge theories with periodic boundary conditions and $2n$ fermionic sites (\zcref{fig:RRW:b}), Gauss's law in the electric basis can be used to construct a \codepar{4n, 2n, 3\!:\!1} (for $n>1$) bitflip code with $H_Z$, the $2n \times 4n$ binary parity check matrix:
\begin{align}
    \scriptsize
    \let\quad\thinspace
    \bordermatrix{
             & S_0 & L_{0,1} & S_1 & L_{1,2} & \ldots & L_{2n-2, 2n-1} & S_{2n-1} & L_{2n-1,0} \cr
    G_0      & 1   & 1       & 0   & 0       & \ldots & 0              & 0        & 1 \cr
    G_1      & 0   & 1       & 1   & 1       & \ldots & 0              & 0        & 0\cr
             &     &         &     &         & \vdots & \cr
    G_{2n-1} & 0   & 0       & 0   & 0       & \ldots & 1              & 1        & 1
    }\,, \label{eq: hz parity check matrix}
\end{align}
\normalsize
where we denoted the Gauss's law check for each site $s$ with $G_s$ (for $n=1$, the code is only an error-detecting \codepar{4,2,2\!:\!1} code).

It is an important implementation detail to note that in our binary mapping of basis states, the syndrome is the same for all the physical basis states, but it is not the trivial syndrome, due to the alternating fermion state mapping. It is common practice in QEC codes to live in a non-trivial coset of the zero syndrome subspace, after a random state preparation procedure. However, Gauss's law QEC coset is inherently deterministic and a result of our staggered fermionic mapping. We can calculate which coset the physical states live in by calculating the syndrome bits for physical states in our qubit mapping corresponding to the even electron sites $G_{2k}$ and the odd positron sites $G_{2k+1}$, $k\in [0,n]$. Electron sites abiding by Gauss's law will have link-site-link configurations in one of $(010),(111)$, $(100), \text{or } (001)$, which all have odd parity for the Gauss's law check described by the corresponding row of $H_Z$ in \zcref{eq: hz parity check matrix}. We note that the last configuration $(001)$ represents $\{-1,e^-,0\}$, which corresponds to a Gauss's law charge sector of $+2$, however, in the 2-modular theories, it is considered to be physical. Positron sites, will be one of $(000)$, $(101)$, $(011)$, or $(110)$, which are of even parity, with $(110)$ ($\{0,e^+,-1\}$, charge sector of -2) similarly only physical in the 2-modular theories. This results in a syndrome of alternating bits for the physical subspace $\boldsymbol{s}_{0}=(1010\ldots01)$. We will, by convention, and for easier mental mapping, transform the measured syndrome by adding $\boldsymbol{s}_{0}$ to it (modulo 2), thus reverting back to the mental model of the physical states being those associated with the zero syndrome. This will ensure that the syndrome of a state corrupted by a bitflip error $\boldsymbol{s}(E_X \ket{\psi})$ will coincide with the syndrome of the error $\boldsymbol{s}(E_X)$.

As a final remark on this affine shift of subspaces, we note that the physical basis states are represented by exactly those bitstrings $\ket{b}$, for which $H_Z \ket{b} =\boldsymbol{s}_0.$ Due to linearity, we can also take any of the physical states, for example, the strong-coupling vacuum of zero particle number $\ket{\Psi_{vac}}$, and shift the kernel of $H_Z$ by the bistring $\boldsymbol{b}_{vac}$ representing $\ket{\Psi_{vac}}$  to find the binary strings representing physical states, $\{\boldsymbol{b} + \boldsymbol{b}_{vac}| \boldsymbol{b} \in \ker{H_Z} \}$. However, elements of $\ker H_Z$ are still important, and will be interpreted as logical transitions between physical states.

To decode errors, a lookup table decoder (\zcref{tab:RRW decoder}) is used to locally decode any single qubit bitflip error using the syndrome of three neighboring lattice sites.
\begin{table}[htbp!]
    \centering
    \begin{tabular}{c|c|c|c}
        $G_{s-1}$ & $G_s$ & $G_{s+1}$ & bitflip     \\
        \hline
        0         & 0     & 0         & none        \\
        1         & 1     & 0         & $L_{s-1,s}$ \\
        0         & 1     & 0         & $S_s$       \\
        0         & 1     & 1         & $L_{s,s+1}$ \\
    \end{tabular}
    \caption{The local lookup table for the RRW decoder for single qubit bitflips. Three lattice site syndromes bits are needed to determine whether the flipped bit is on the center site or the incoming or outgoing link. The syndrome bits should be interpreted as already being offset by $\boldsymbol{s}_0$.}
    \label{tab:RRW decoder}
\end{table}
To correct all errors, not just bitflip errors, we can concatenate the RRW code on top of a $d=3$ phase-flip code. This results in weight-9 stabilizer generators, as displayed in \zcref{fig:RRW:c} for an $n=2$ lattice.

The RRW code is clearly able to save on the number of qubits compared to using Shor's code for each qubit as in \zcref{fig:RRW:a}. However, as we will see later, these savings come with a price, as this construction is incompatible with the non-periodic $U(1)$ theories and has an increased mixing speed penalty.

\section{Gauss's law stabilizer codes require a periodic electric field} \label{sec: dim of schwinger}

For our first result, we show precisely why Gauss's law error correction using stabilizer codes limits the lattice gauge theory design space to periodic electric fields.

The RRW method measures the eigenvalues of the $G_i$ Gauss's law checks using a qubit, thus the measured charge sector is modulo 2. Similarly, the $U(1)_2^{\circ}$ (and the $\mathbb{Z}_2$) theories use a mod 2 version of Gauss's law. Thus, intuitively it is clear that the $U(1)_2^{\circ}$ (and the $\mathbb{Z}_2$) massive Schwinger models are compatible with RRW, namely, the code space of RRW coincides with $\mathcal{H}_{phys}^{U(1)_2^{\circ}}=\mathcal{H}_{phys}^{\mathbb{Z}_2}$, as was shown in \cite{rajput_quantum_2023,spagnoli_fault-tolerant_2024}. However, this is not the case for the non-periodic theory, as $\mathcal{H}_{phys}^{U(1)_2^{-}} \subset \mathcal{H}_{phys}^{U(1)_2^{\circ}}$. One might hope to construct an alternative stabilizer code with a smaller codespace that accommodates the non-periodic electric-field theories. We will show that this is not possible, by showing that $\dim\mathcal{H}_{phys}^{U(1)_2^{-}}(n) = \mathcal{L}(2n)$, where $\mathcal{L}(k)$ is the $k^{\text{th}}$ Lucas number. Only two of the smallest Lucas numbers are perfect powers. Thus, they cannot be the dimension of the code space of a stabilizer code.

In the following, we directly calculate $\dim\mathcal{H}_{phys}^{U(1)_2^{-}}(n)$ by building a graph representation and path counting strategy. Denote the flux value on the incoming $L_{s,s-1}$ and the outgoing $L_{s+1,s}$ links with a pair $(i_s,o_s)$, the \textit{flux configuration}. The possible configurations $(i_s,o_s)$ are limited by the finite truncation, and because the charge $\rho_s=o_s-i_s$ is -1 or 0 for electrons and +1 or 0 for positrons. In the $U(1)_2^{-}$ theory on a site $s$, Gauss's law satisfying configurations include: $\mathcal{C}_{e^-}=\{(-,-),(0,-),(0,0)\}$ for electron sites and $\mathcal{C}_{e^+}=\{(-,-),(-,0),(0,0)\}$ for positron sites. Next, we create a graph where paths will represent the possible states of three fermion sites ($n_f = 3$) with open boundary conditions, which provides an extendable unit that can be wrapped into periodic boundary conditions.

\begin{figure}
    \centering
    \vspace{5px}
    \includegraphics[width=1\linewidth]{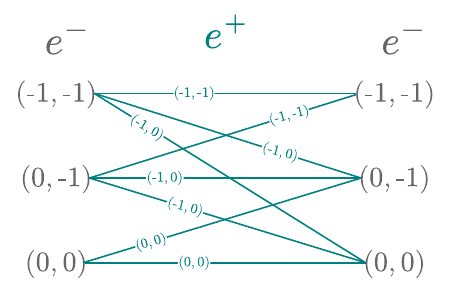}
    \vspace{2px}
    \caption{The flux configuration space graph between two neighboring electron sites for a $U(1)_2^-$ theory. Notably, the graph is almost an all-to-all graph, except for one edge corresponding $(0,-1)$, which is not a physical flux configuration for positron sites.}
    \label{fig:flux config space graph}
\end{figure}

Let the flux configuration space graph be such that nodes are possible flux configurations around electron sites. Two electron configurations $(i_s, o_s)$ and $(i_{s+2}, o_{s+2})$ are connected, when there exists a corresponding positron type configuration $(o_{s},i_{s+2}) \in \mathcal{C}_{e^+}$. For a visual example, see \zcref{fig:flux config space graph}.

We order the configurations alphanumerically, and then, we can create an adjacency matrix of this graph:
\begin{align}
    A = \bordermatrix{
    s+2 \backslash s & (-,-) & (0,-) & (0,0) \cr
    (-,-)            & 1     & 1     & 0  \cr
    (0,-)            & 1     & 1     & 1  \cr
    (0,0)            & 1     & 1     & 1
    } \label{eq: config counting matrix for d2} \,.
\end{align}
For periodic boundary conditions, physical states correspond to paths that start and end in the same flux configuration. We can count these by taking the trace of the adjacency matrix. For $n$ physical sites, the graph can be described with $n$ copies of the adjacency matrix $A$, and the number of physical states is $\tr[A^n]$. This is just the sum of the eigenvalues $\lambda_i^n$ of $A^n$. The eigenvalues of $A$ are $\lambda_0=0, \lambda_+ = \frac{3+\sqrt{5}}{2}, \lambda_- = \frac{3-\sqrt{5}}{2}$, thus:
\begin{align}
    \dim \mathcal{H}_{phys}^{U(1)_2^-} (n) & = \tr[A^n] = \sum_{i} \lambda_i^n \label{eq: dim is sum of eigvals} \\
                                           & = \frac{(3+\sqrt{5})^n + (3-\sqrt{5})^n}{2^n}\,.
\end{align}
Curiously, this is exactly the bisection of the Lucas numbers \footnote{{https://oeis.org/A005248}}, $\mathcal{L}(2n)$, with the Lucas numbers recursively defined as follows:
\begin{align}
    \mathcal{L}(i)=\begin{cases}2, & i=0 \\ 1 , &i=1 \\ \mathcal{L}(i-1) + \mathcal{L}(i-2)
                & i>1\end{cases}\ \ \ .
\end{align}
This means that for $U(1)_2^-$ a Gauss's law QEC code should have $\mathcal{L}(2n)$ dimensional logical basis states. For stabilizer qubit codes, the code space dimension is $2^k$. As the only Lucas numbers that are perfect powers are $\mathcal{L}(2)=1, \mathcal{L}(4)={4}$ \cite{bugeaud_classical_2006}, this means that no stabilizer code family matches this dimensionality requirement. Furthermore, the gap in physical dimension between the periodic and non-periodic electric field theories is exponentially large,
\begin{equation}
    \begin{aligned}
        \lim_{n \rightarrow \infty} \frac{\dim \mathcal{H}_{phys}^{U(1)_2^{\circ}} }{ \dim \mathcal{H}_{phys}^{U(1)_2^{-}}} & = \lim_{n \rightarrow \infty} \frac{2^{3n}}{(3+\sqrt{5})^n + (3-\sqrt{5})^n} \\
                                                                                                                            & \approx \lim_{n \rightarrow \infty} 1.53^n  \ \ .
    \end{aligned}
\end{equation}
We conjecture that similar results will hold for gauge field dimensions $d>2$, based on thoughts in \zcref{app: higher dimensional counting}.

\section{Simulation results} \label{sec: simulations}

Our second contribution is the benchmarking of the RRW code relative to an application-agnostic, universal quantum error correcting code in the important context of simulation performance.

\subsection{Methods}

In the following section, simulated systems focus on $n = 2$ site $U(1)_2^{\circ/-}$ lattice gauge theories (see \zcref{app: modular} for $\mathbb{Z}_2$). Our choice of universal QEC scheme is the simple $d=3$ bitflip code, which, concatenated similarly to the RRW code, results in the \codepar{9,1,3} X-Shor codes. See \zcref{fig:RRW:a} for a layout of the stabilizer generators.

We hone in on the bitflip axis of error correction, hence our noise model is simply the independent bitflip noise. The small, 8-qubit size of the lattices allows our simulations to leverage the density matrix formalism and Kraus operators using QuTIP 5 \cite{lambert_qutip_2024, johansson_qutip_2013, johansson_qutip_2012}.

\subsubsection{Decoding methods} \label{sssec: decoding methods}

To decode the states corrupted by independent, identically distributed, probabilistic bitflips, we implement two different minimum weight decoding strategies. One is minimum weight perfect matching (MWPM) using PyMatching \cite{higgott_sparse_2025}, the second is using runs of ones \cite{moodDistributionTheoryRuns1940} for analytical calculations. Next, we explain our motivation for the two alternative methods and their equivalence to the RRW solution in the specific context of single-qubit bitflips.

The original RRW lookup-table solution only works for single-qubit bitflips and multiple bitflips that are well-separated. One can see this from the fact that for three bits corresponding to three neighboring sites, only four of the eight possible bitstrings are captured in \zcref{tab:RRW decoder}. While it is impossible to correct all multi-qubit errors, a decoder can make a bet on recovery for all syndromes, thus increasing the chance of recovery and improving the overall logical error rate. Thus, the RRW decoder needs to be extended to a minimum-weight decoder to cover all the multi-qubit cases. This is not hard, though it is already subtle in our 1+1D case, as is evident in our \textbf{extended RRW decoder} described in \zcref{app: extended lookup table decoder}. Decoding is typically more complex for lattices with higher spatial dimensions. A natural alternative that we recommend considering is to use the minimum-weight perfect matching algorithm.

In the parity check matrix $H_Z$ in \zcref{eq: hz parity check matrix}, all bitflips only impact at most two Gauss's law checks. This makes the detector error model \cite{derksDesigningFaulttolerantCircuits2025} graph-like, as at most two detectors (checks) are activated by a single bit flip. Decoding syndromes based on graph-like error models is possible by efficiently solving the minimum-weight perfect matching problem on the violated checks. This method generalizes to higher-dimensional lattice gauge theories as well; in 2D+1 and 3D+1 models, two Gauss's law checks will still overlap on one link qudit, creating a graph-like detector error model. MWPM/PyMatching decodes syndromes for more than one bitflips deterministically with a minimum weight solution. Given MWPM/PyMatching's unified decoding across 1+1D, 2+1D and 3+1D and ease of implementation, and given the well-tested PyMatching library, MWPM is an attractive choice for decoding Gauss's law QEC.

For analytical calculations, we analyze runs of ones (or simply ``runs'' are substrings of ones, flanked by zeros on each end) \cite{moodDistributionTheoryRuns1940} in syndrome strings with periodic boundary conditions. We notice from the lookup table in \zcref{tab:RRW decoder} that bitflips on site qubits generate a single syndrome bit, and link qubit bitflips generate two syndrome bits. For a minimum-weight decoding scheme using the RRW lookup table, we observe that each run of ones of length $\ell_j$ is correctable by a combination of $\lceil\frac{\ell_j}{2}\big\rceil$ link and site qubits. This is because even-length runs $2k$ can be recovered using the $k$ link qubits between them, and odd-length runs $2k+1$ can be turned into an even length with one extra site qubit. Thus, the minimum weight of the recovery operator $R(\boldsymbol{s})$ for a syndrome $\boldsymbol{s}$ with a \textit{vector of run lengths} $\boldsymbol{\ell}$ can be rapidly calculated as
\begin{align}
    |R(\boldsymbol{s})| = \sum_{\ell_j \in \boldsymbol{\ell}} \big\lceil\frac{\ell_j}{2}\big\rceil \,. \label{eq: runs of ones min weight}
\end{align}

\subsubsection{Logical channels} \label{sssec: channels}

We define the logical channel as the effective channel that describes the evolution of the logical states. This is the combination of the physical noise, error correction recovery, and potentially Hamiltonian evolution channels. When using no error correction (no QEC), this is just the physical channel,
\begin{align}
    \mathcal{E}^{\text{noQEC}}_p(\rho) = \sum_i A_i^{\text{noQEC}}(p) \rho \bigl( A_i^{\text{noQEC}}(p) \bigr)^\dagger\,,
\end{align}
with the following $2^N$ Kraus operators for $N=4n$ qubits:
\begin{equation}
    A_{\mathbf{v}}^{\text{noQEC}}(p) = \sqrt{p^{|\mathbf{v}|}(1-p)^{N-|\mathbf{v}|}} \prod_{j = 0}^{N-1} X_{j}^{v_j}  \label{eq: no qec kraus} \,,
\end{equation}
where the vectors $\mathbf{v}$ are binary vectors of length $N$ with $|\mathbf{v}|$ (the Hamming weight of $\mathbf{v}$) the number of 1s designating error locations. When using universal QEC (UQEC), the $p$ parameter is simply scaled to be the logical error rate of the $d=3$ bitflip code, $p_3=3p^2(1-p)+p^3$. Thus, the logical channel for UQEC is,
\begin{align}
    \mathcal{E}^{\text{UQEC}}_p(\rho)=\mathcal{E}^{\text{noQEC}}_{p_3}(\rho) \,.
\end{align}
For the Gauss's law QEC (GLQEC) logical error channel $\mathcal{E}^{\rm GLQEC}_p(\rho)$, the Kraus operators from \zcref{eq: no qec kraus} are simply modified by a factor of a recovery operator $R(\boldsymbol{s}(E_\mathbf{v}))$ that we find through decoding:
\begin{align}
    A_{\mathbf{v}}^{\text{GLQEC}}(p) =R(\boldsymbol{s}(E_{\mathbf{v}})) A_{\mathbf{v}}^{\text{noQEC}}(p) \,, \label{eq: glqec kraus}
\end{align}
where $E_{\mathbf{v}}=\prod_{j = 0}^{N-1} X_{j}^{v_j}$ is the same error operator corresponding to $A_{\mathbf{v}}^{\text{noQEC}}$. In general, there will be redundancies in the expression of \zcref{eq: glqec kraus} as $|\mathbf{v}| = 2^{4n}$ while there are only $2^{2n}$ unique $A^{\text{GLQEC}}$ operators. Due to linearity, note that this modification of Kraus operators can equivalently be implemented by folding the density matrix $\mathcal{E}_p^{\rm noQEC}(\rho)$ into the code space via recovery of the pair of basis states addressing the location of each matrix element.

The above channels are commonly referred to as code capacity level noise models because we assume perfect syndrome extraction, gates, and measurements. Functionally, these three channels are also the underlying mechanism for \textit{memory experiments}, as the evolution is the logical identity operator.
For \textit{simulation experiments}, for every time step $\delta t=1/3$, we unitarily evolve the system with $e^{-i\delta tH}$ and then apply one of the noisy channels above. These code capacity simulations allow us to isolate the effects of the error-correction schemes at the most fundamental level, without other sources of errors, such as circuit-level noise or Trotterization errors.

\subsection{Results}

\subsubsection{Single-round memory experiment}\label{ssec:single-round memory}

To benchmark QEC codes under code capacity noise models, it is typical to run single-round memory experiments. Here, the effect of noise on the data qubits is corrected in a single QEC round. Then, if the resulting state is the same logical codeword as the initial state, no logical error has occurred; otherwise, a logical error has occurred. In a typical Monte Carlo experiment, each sample is repeated, and the average logical error rate is calculated. Here, instead, we take an analytical approach for the noQEC and the UQEC channels, and we turn to combinatorics by counting runs of ones in all possible syndrome strings for the GLQEC channel.

\begin{figure}[!htbp]
    \includegraphics[width = 0.45\textwidth]{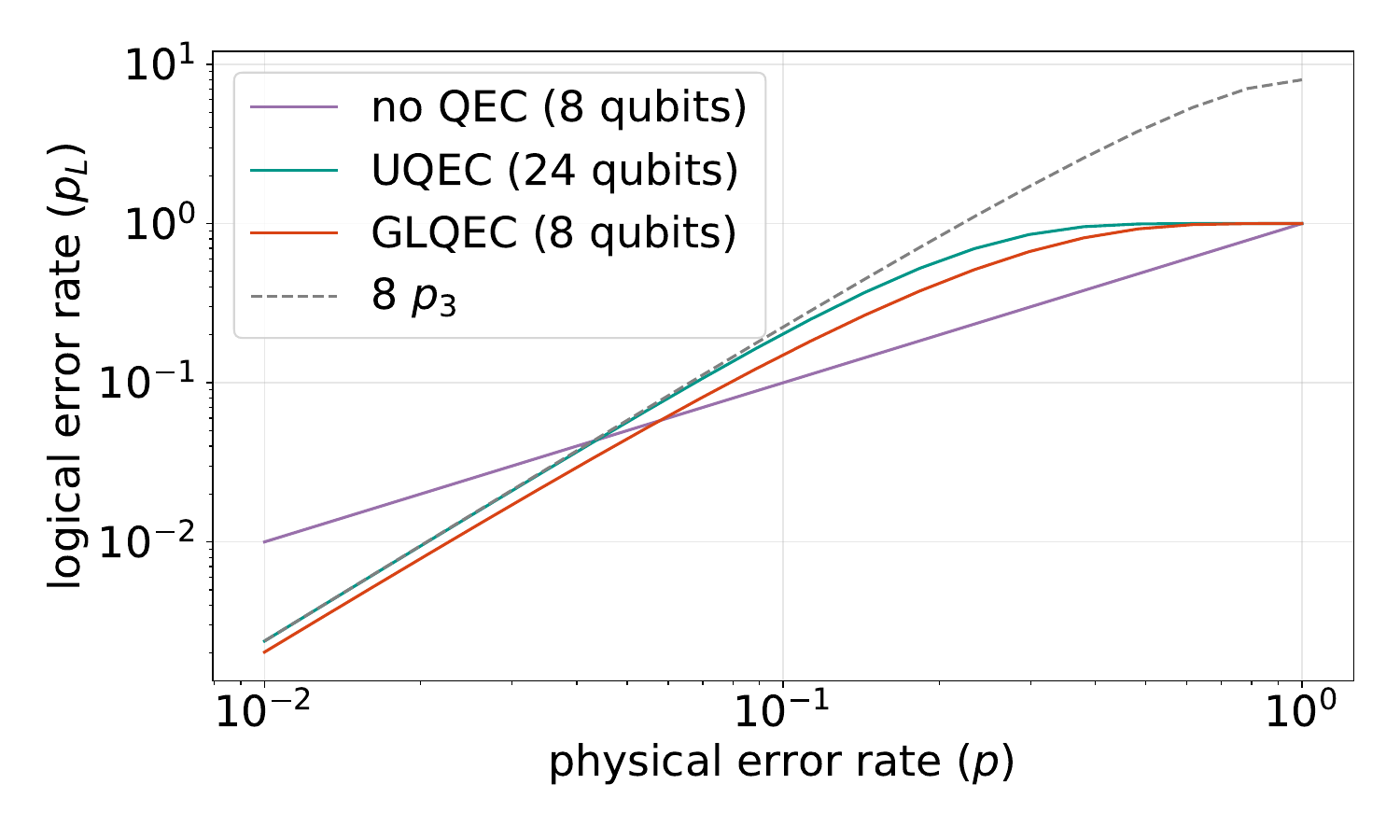}
    \caption{Logical error rates as a function of physical error rate for a lattice of $n=2$ physical sites subject to universal (UQEC) or Gauss's law error correction (GLQEC) strategies compared to the bare physical qubit performance (noQEC). The noise channel is bitflip noise with probability $p$, and for simplicity, the codes are not concatenated with the phase-flip code (as opposed to \zcref{fig:rrw lattice}). As expected for two $d=3$ codes, GLQEC achieves the same $p^2$ order of asymptotic scaling with fewer qubits using 3-body stabilizers, compared to the 2-body stabilizers used in UQEC.}
    \label{fig:memoryErrorRate}
\end{figure}

\begin{figure}[!htbp]
    \includegraphics[width =1.1\linewidth]{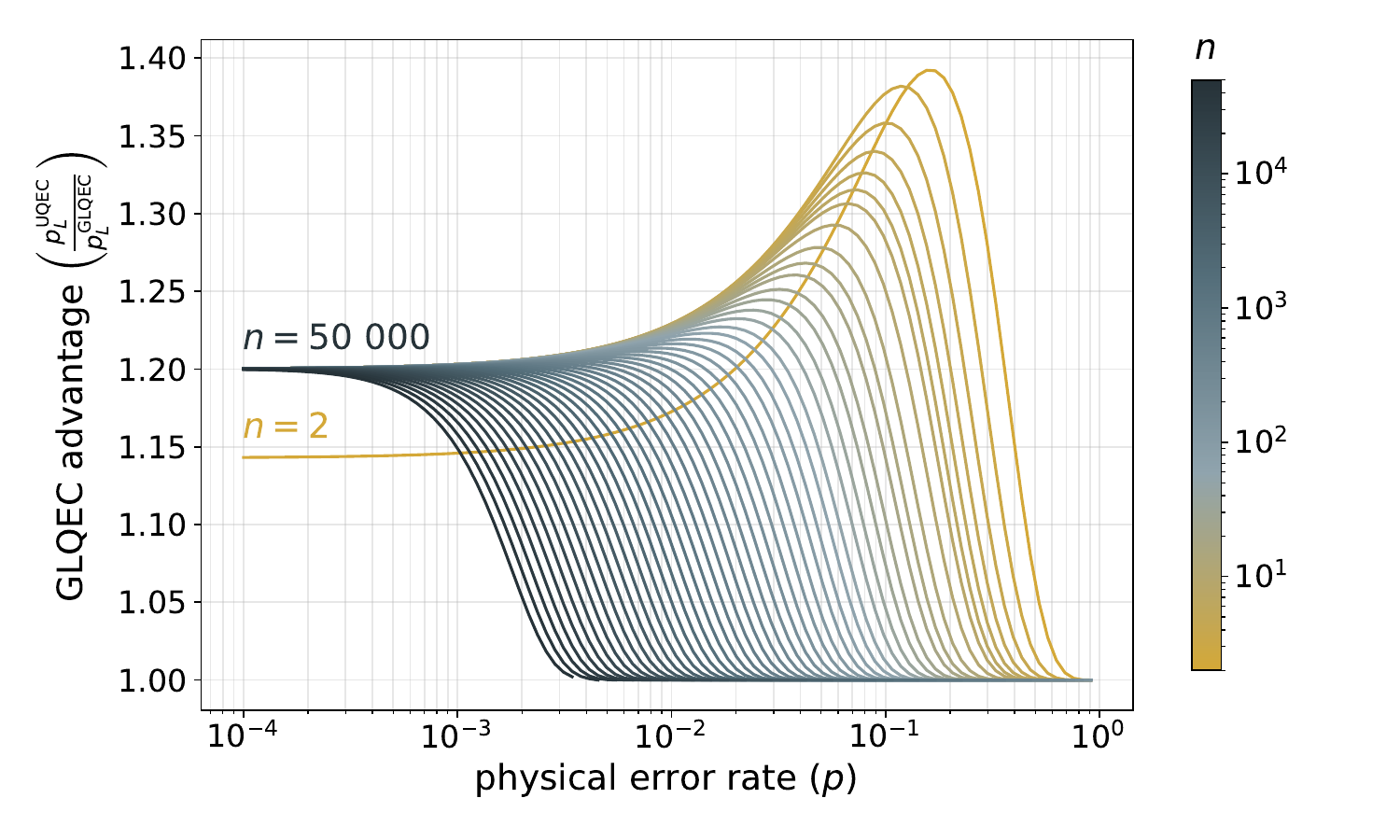}
    \caption{Ratio of universal to Gauss's law logical error rates as a function of physical error rate for lattices of increasing volume from $n = 2$--$50\ 000$ physical sites.}
    \label{fig:errRatio}
\end{figure}

With universal error correction of distance three, the error rate experienced by a lattice of $n$ physical sites becomes,
\begin{align}
    p_L^{\text{UQEC}} & = \sum_{k = 1}^{4n} \binom{4n}{k} p_3^k(1-p_3)^{4n-k} \label{eq:univErrRate} \\
                      & =1-(1-p_3)^{4n}\,.
\end{align}
As shown in \zcref{fig:memoryErrorRate}, this use of extra physical qubits provides a logical error rate approaching $4np_3$ at small physical error rates.

In order to derive the small-$p$ behavior of the Gauss's law error correction scheme, we leverage \zcref{eq: runs of ones min weight} to calculate the minimum weight of the recovery for a syndrome.
For a lattice of $n$ physical sites, let $C_{n,k}$ denote the number of correctable weight-$k$ bitflip errors, with $\sum_k C_{n,k} = 2^{2n}$. $C_{n,k}$ can be formulated combinatorially as the number of runs of ones configurations under periodic boundary conditions in $2n$ bits corresponding to a minimum-weight recovery of weight $k$. More precisely, $C_{n,k}$ is the number of all runs of ones configurations with a vector of run lengths $\boldsymbol{\ell}$ that satisfies
\begin{align}
    \sum_j \big\lceil\frac{\ell_j}{2} \big\rceil                                     & =k \text{, with}                                    \\
    \sum_j\ell_j + \left(1-\delta_{1,|\boldsymbol{\ell}|}\right) |\boldsymbol{\ell}| & \leq 2n \label{eq: runs of ones qubit bound}\ \ \ ,
\end{align}
where $|\boldsymbol{\ell}|$ is the number of runs, and \zcref{eq: runs of ones qubit bound} constrains the total number of ones plus the separating zeros to be less than or equal to the number of syndrome bits. From this, we can see that $k\leq n$, because the highest weight syndrome is the syndrome with $2n$ ones, which will be corrected to the weight $n$. Thus,
\begin{align}
    p_L^{\rm GLQEC}(p) =1- \sum_{k=0}^n C_{n,k} p^k(1-p)^{4n-k}
\end{align}
is the GLQEC logical error rate for the single-round memory experiment.

By analyzing a finite state machine that can keep track of the $\sum_j\lceil \frac{\ell_j}{2}\rceil$ statistic in syndromes, we derive a generating function $W_n(u) = \sum_{k=0}^n C_{n,k} u^k$ in \zcref{app: analytical runs of 1s count},
\begin{align}
    W_n(u)
     & = \left(\frac{1-\sqrt{1+8u}}{2}\right)^{2n} + \nonumber \\& \qquad \left(\frac{1+ \sqrt{1+8u}}{2}\right)^{2n} - u^n \label{eq: gen func}\ ,
\end{align}
leading to an analytic expression for its coefficients $C_{n,k}$,
\begin{align}
    C_{n,k} & = 2^{3k-2n+1}{\sum_{m=k}^n \binom{2n}{2m} \binom{m}{k}}-\delta_{n,k} \label{eq: cnk} \ .
\end{align}
Thus, with $p^k(1-p)^{4n-k} = (1-p)^{4n}\left(\frac{p}{1-p}\right)^k $, we can calculate the GLQEC error probability using the generating function in \zcref{eq: gen func}:
\begin{equation}
    p_L^{\rm GLQEC}(p)=1-(1-p)^{4n}W_n\left(\frac{p}{1-p}\right) \label{eq: pl glqec analytical}\,.
\end{equation}
Using \zcref{eq: pl glqec analytical}, we show the GLQEC channel's advantage over UQEC, $\frac{p_L^{\rm UQEC}(p)}{p_L^{\rm GLQEC}(p)}$, in \zcref{fig:memoryErrorRate} from $n=2$ up to $n=50\ 000$. We observe that the advantage curve for $n=2$ physical sites differs from that of larger systems due to finite-size effects. For large $n$, GLQEC slightly outperforms the universal error correction scheme up to a constant factor of $1.2$ in the low-$p$ regime. The high-$p$ regime shows no GLQEC advantage, as evidenced by a ratio of 1.0. The mid-$p$ regime shows a transition between $1.2$ and $1.0$, with a peak present only for small system sizes. In the plot, the system sizes are evenly spaced in log space from $n=2$ to $n=50\ 000$. The even spacing observed at a constant GLQEC advantage (for example, at $\frac{p_L^{\rm UQEC}(p)}{p_L^{\rm GLQEC}(p)}=1.05$) indicates that the system size at which the GLQEC advantage goes below this level grows polynomially with decreasing physical error rate. For example, at $p=0.002$ the GLQEC advantage remains above 1.05 for lattice volumes below $n\approx50\ 000$.

To understand the 1.2 low-$p$ ratio, we can focus on syndrome sequences decoded to two-qubit errors. There are five distinct $k=2$ runs of ones configurations $\boldsymbol{\ell}$: $(3), (4), (1,1), (1,2)$, and $(2,2)$.
The total number of these runs of ones configurations yields the number of correctable syndrome configurations at order $p^2$,
\begin{equation}
    \begin{aligned}
        C_{n,2} & = 2n(1-\delta_{n,1})                      \\
                & + (1-\delta_{n,1})(2n+(1-2n)\delta_{n,2}) \\
                & +\max[n(2n-3),0]                          \\
                & +\max[2n(2n-4),0]                         \\
                & +\max[n(2n-5),0] \,,
    \end{aligned}
\end{equation}
where each line corresponds to the contribution from the five runs of ones configurations above. At $n\geq3$, the number of incorrectable errors at order $p^2$ is $10n$ from
\begin{equation}
    C_{n,2} = \binom{4n}{2} - 10n\text{, when }n\geq 3\,. \label{eq: num correctable weight-2}
\end{equation}
Comparing this with the incorrectable errors at order $p^2$ under universal error correction, $3\binom{4n}{1}$ via \zcref{eq:univErrRate}, yields a $\frac{6}{5}$ ratio of improvement at small-$p$  as seen in \zcref{fig:errRatio}.
For physical error rates $\sim10^{-4}$, this asymptotic improvement in the logical error rate is achieved for lattice volumes $50\ 000 \gtrsim n \geq 3$. However, because the constant distance $d=3$ holds for both codes, we expect the advantage to disappear for any fixed-$p$ at asymptotic system sizes,
\begin{align}
    \lim_{n \rightarrow \infty}\frac{p_L^{\rm UQEC}(p)}{p_L^{\rm GLQEC}(p)} = \frac{1}{1}= 1\,.
\end{align}
since the number of possible errors grows in both systems.

\subsubsection{Noisy evolution}\label{ssec:noise evolution}

To further evaluate the performance of the Gauss's law QEC solution compared to universal error correction, we benchmark the following observables of interest throughout noisy evolution in truncated $U(1)$ theories $U(1)_2^\circ$ and $U(1)_2^-$:

\begin{itemize}
    \item $\langle e^+e^- \rangle$ is the probability of being in a (not necessarily physical) state with exactly one pair of $e^-$ and $e^+$, thus the expectation value of the sum of projectors for all states with a single pair, $\text{Tr}[\Pi_{e^+e^-} \rho_t]$.
    \item $\langle H_E \rangle$ is the expectation value of the energy in the electric field, $\text{Tr}\left[ H_E \rho_t\right]$ from \zcref{eq: hamiltonian}.

    \item $\bra{\Psi_{vac}}\rho_t\ket{\Psi_{vac}}$ is the overlap with the strong coupling vacuum state $\ket{\Psi_{vac}}$.
    \item $\langle \Pi_{G=0}\rangle$ is the probability of being in the physical subspace, thus the expectation value of the sum of projectors for all physical basis states. For a given theory, its own Gauss's law is used, namely, a non-periodic Gauss's law is used for $U(1)_d^{-}$ and a periodic Gauss's law is used modulo $d$ for $U(1)_d^{\circ}$.
\end{itemize}

\begin{figure*}
    \includegraphics[width=1\textwidth]{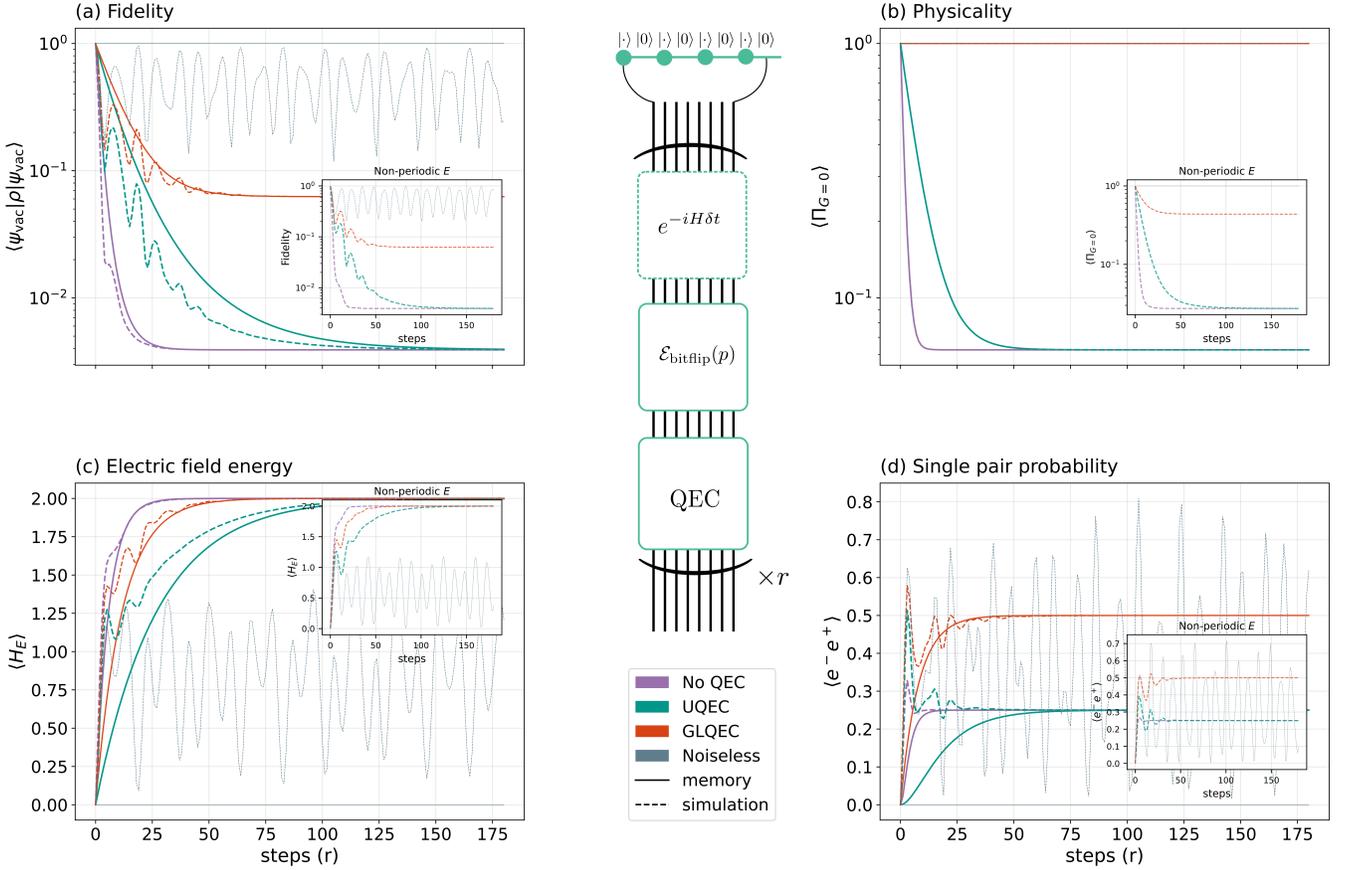}
    \caption{Multi-round memory and simulation results for the truncated $U(1)$ theories for $n=2$ lattice size (8 application qubits) and $p=0.08$ error rate. Exact, density-matrix-based expectation values for fidelity (a), physicality (b), electric field energy (c), and single pair probability (d) are shown for the cases of no error correction (no QEC), universal error correction (UQEC) with the bitflip code, and Gauss's law error correction (GLQEC). As a baseline, the noiseless cases are displayed in gray. All series are shown for memory and simulation channels for the $U(1)_2^{\circ}$ theory. The insets show the highly similar results for $U(1)_2^{-}$ theories.}
    \label{fig: combined multi round}
\end{figure*}

Using density matrix formalism and the channels defined in \zcref{sssec: channels}, starting from the strong-coupling vacuum state of zero particles, we perform time evolution for 180 steps with $\delta t=\frac{1}{3}$ and evaluate the observables above for memory as well as Hamiltonian evolution cases, based on \zcref{eq: hamiltonian} with parameters $x=0.6, \mu=0.1$ using the qubit mapping of the site and link operators defined in \zcref{ssec: qubitized lgt}. We plot the results combined in \zcref{fig: combined multi round}.

A clear advantage of using GLQEC arises in the physicality for the GLQEC-compatible $U(1)^\circ_d$ theory. In \zcref{fig: combined multi round} (b), we can see that the theory with the periodic electric field maintains perfect physicality. However, the non-periodic theory still faces a drop in physicality when using GLQEC, as GLQEC's logical subspace contains extra states that are only modulo 2 physical, as discussed in \zcref{sec: dim of schwinger}. Though the current analysis has not been optimized for this purpose, it is interesting to note from the insets that the presence of unphysical states in the logical codespace does not appear to significantly degrade simulation quality, warranting more dedicated analysis if desired in future simulation designs.

Despite maintained physicality, lower qubit usage, and better single-round memory performance, we observe the surprising effect that the GLQEC experiments decohere faster to the completely mixed state than the UQEC experiments. This behavior is also evident in multi-round memory experiments without the Hamiltonian evolution.

Prior to thermalization, there are cases in which the GLQEC expectation value is closer than its UQEC counterpart to the noiseless expectation value. This, however, should not be interpreted as superior performance of GLQEC relative to UQEC, as it can be attributed to differences in the final asymptotic expectation values between the two channels. This effect is inconsistent across different observables, as seen in \zcref{fig: combined multi round}. For fidelity/electric energy, GLQEC/UQEC is closer to the noiseless case, whereas the two are nearly equidistant from the mean noiseless value for the single fermion pair probability.

\section{Mixing speed analysis}\label{ssec:mixing speed analysis}

It is surprising that, although GLQEC outperforms UQEC in a single-round memory experiment, it exhibits faster decoherence over time, making its relative advantage highly sensitive to its thermal value in the codespace. Here, we seek to further our understanding of the reasons for this behavior. We limit our analysis to the binary truncated $U(1)$ theory with a periodic electric field $U(1)^{\circ}_2$, as it is compatible with GLQEC and the mixing behavior is similar in all theories. For this exploration, we will focus on multi-round memory experiments, rather than Hamiltonian evolution, as they capture the essential mixing dynamics. They also allow a significantly simpler analysis based on Markov processes, given their essentially classical behavior. We are especially interested in the ranking of mixing times as the system size grows. We present further numerical evidence that the ranking is preserved up to 50 physical sites (200 application qubits), and numerically determine a mixing threshold for GLQEC. Finally, we analytically calculate the second largest eigenvalue modulus (SLEM) of the logical channels under a truncated Markov model.

\subsection{Multi-round memory Monte Carlo}

\begin{figure}[htbp!]
    \centering
    \includegraphics[width=1\linewidth]{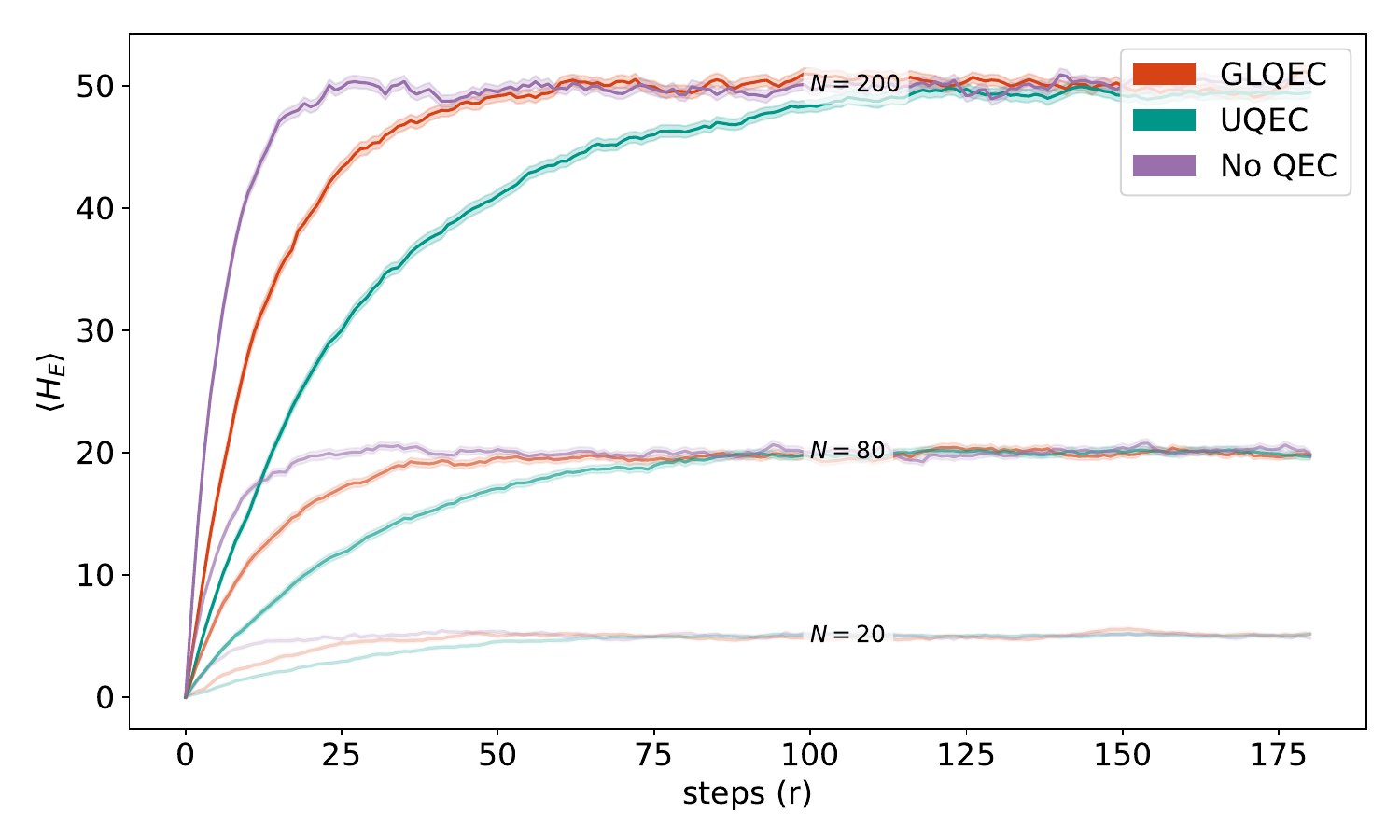}
    \caption{Monte Carlo multi-round memory simulation of $\langle H_E \rangle$ at $p=0.08$ error rate for $n=5,20$ and 50 sites (corresponding to $N=20,80$ and 200 application qubits respectively) with sample size 100 for each time step. Shading is the standard deviation of the samples at each time step.}
    \label{fig:e2 for n20}
\end{figure}

As shown in \zcref{sec: simulations} for $n=2$ physical sites, the mixing speed of the GLQEC channel is between that of the universal and the noQEC channels. Central questions following this observation include: whether this ranking of channel mixing times is preserved as system size increases, how it depends on the physical noise strength $p$, and what role finite-size effects play. As we are only focusing on a memory channel, we can use a classical bitstring to represent the state of the system along a single probabilistic trajectory and use Monte Carlo sampling to explore significantly larger systems than would be possible with the computation-intensive density matrix formalism. While calculating the trace distance would require effectively reassembling the density matrix from the state vector samples, we instead use the expectation value of $H_E$, which has the same thermal expectation value across all three channels, as a simpler proxy for mixing. In \zcref{fig:e2 for n20}, we see that at $n=5,20$, and 50 sites (corresponding to $n_q=20,80$, and 200 qubits respectively) there is evidence that the ranking is preserved in larger systems. The presented thermalizations start from the strong coupling vacuum state, though we expect the decoherence to be insensitive to the initial state because the bitflip channel is state independent. The curves resemble those in the $n=2$ case, suggesting that finite-size effects are not significant. Next, we develop an analytic approach to this problem using spectral analysis.

\subsection{Second largest eigenvalue modulus analysis} \label{ssec: mixing speed analysis for lambda2}

All three channels considered in this work are completely positive and trace-preserving, and thus can be thought of as quantum Markov channels \cite{terhalProblemEquilibrationComputation2000}. Due to the effect of the bitflip channel, the steady-state for each channel is the completely mixed state $\rho \propto I_C$ within a channel-dependent subspace $C$. The subspace $C$ can be one of $C^\text{noQEC},C^\text{UQEC}$ or $C^\text{GLQEC}$, corresponding to our three channels.

For the channel without error correction, $I_C$ is the mixed state within the physical Hilbert space of the $4n$ qubits. Thus, using $\simeq$ to denote isomorphism between two vector spaces, $C^{\text{noQEC}}=\mathcal{H} \simeq \mathbb{C}^{\otimes 4n}$. Similarly, for the universal error correction $C$ is the code space of $4n$ logical qubits $C^{\text{UQEC}} \simeq C^{\text{noQEC}} \simeq \mathbb{C}^{\otimes 4n}$. For GLQEC, the evolution is constrained to the code space of the $2^{2n}$ basis states that are physical with respect to the modular Gauss's law, $C^{\text{GLQEC}} = \mathcal{H}_{phys}^{U(1)_2^\circ} \simeq \mathbb{C}^{\otimes 2n}$.

\begin{figure}[htbp!]
    \centering
    \includegraphics[width=1\linewidth]{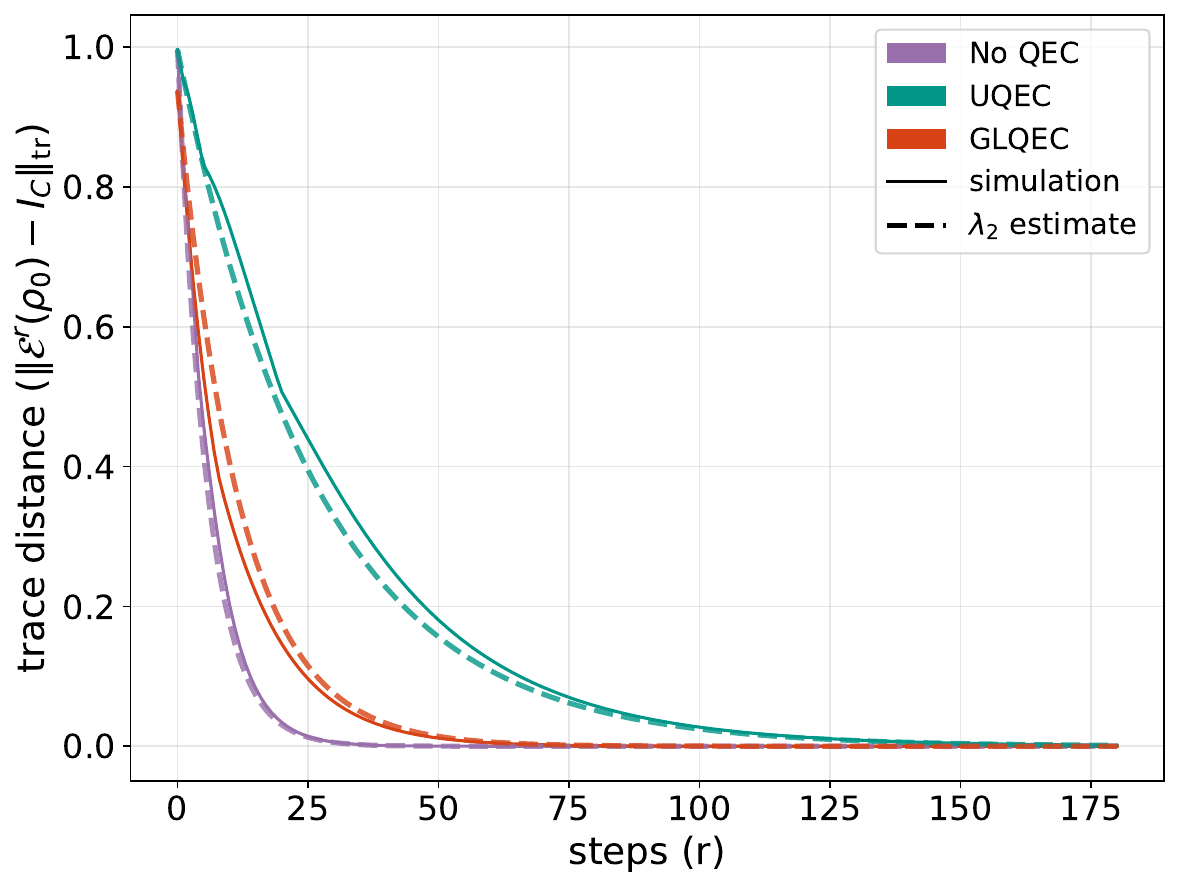}
    \caption[Trace distance and its SLEM-based estimators.]{Trace distance from the mixed states for $n=2$ density matrix simulations at $p = 0.08$ error rate, including the $\lambda_2$-based estimators.}
    \label{fig:slem}
\end{figure}

The trace distance of two density matrices $\sigma, \rho$ is defined as half of the trace norm of the difference between the two matrices
\begin{align}
    |\!|\sigma - \rho|\!|_{\rm tr}=\frac{1}{2} \tr[\sqrt{(\sigma-\rho)^\dagger(\sigma-\rho)}]\,.
\end{align}
To study how fast the system reaches $I_C$, it is possible to provide upper bounds on the trace distance between the state $\mathcal{E}^r(\rho)$ at step $r$ and the final mixed state $I_C$ by using the second largest eigenvalue modulus of the quantum channel \cite{temme_2-divergence_2010, terhalProblemEquilibrationComputation2000}. The key idea is that
\begin{align}
    |\!|\mathcal{E}^r(\rho_0) - I_C|\!|_{\rm tr} \leq D(\rho_0,I_C) |\lambda_2|^r\, ,
    \label{eq:slemBound}
\end{align}
for some constant $D(\rho_0,I_C)$ where $|\lambda_2|$ is the SLEM of the channel $\mathcal{E}$. In Ref.~\cite{temme_2-divergence_2010}, $D(\rho_0,I_C)=\sqrt{\chi^2(\rho_0,I_C)}$ is the $\chi^2$-divergence. We find that for $D(\rho_0,I_C)=|\!|\rho_0 - I_C|\!|_{\rm tr}$, the right side of \zcref{eq:slemBound} is not an upper bound, but actually works well as an estimator for our two logical channels and the physical bitflip channel. In \zcref{fig:slem}, we see that our $\lambda_2$-based estimators closely agree with the trace distance for $n=2$ density-matrix simulations.

Motivated by this estimator for mixing rate, we next present the calculation of $\lambda_2$. As our memory channels are essentially classical, we can estimate $\lambda_2$ using classical Markov chain analysis. While it is well-known that the bitflip channel contracts the $Y$ and $Z$ components of the Bloch sphere by a factor of $1-2p$ \cite{nielsenQuantumComputationQuantum2010a}, we derive this result for $\lambda_2$ as a way of also introducing the Fourier analysis for Markov chains, which we will utilize later for the analysis of the GLQEC channel.

Both the classical bitflip channel on $4n$ bits and the quantum bitflip channel on $4n$ qubits are discrete convolutions on the Abelian group $G=\mathbb{Z}_2^{\otimes 4n}$ consisting of binary strings of length $4n$, which is closed under addition modulo 2. The bitstrings can be viewed as representations of the transitions (or jumps, or differences) $\tau_{ij}=\tau_{ji}=b_i + b_j$ between the basis states $b_i,b_j$ in $C$. Given an ordering of the basis states in the logical channel with symmetric transition probabilities $p(\tau_{ij} )=p(\tau_{ji} )$, we can define the discrete convolution corresponding to the Markov chain $M$ on a function $f: G \rightarrow A$ with arbitrary codomain $A$ as
\begin{align}
    (Mf)(x) = \sum_{\tau_{ij} \in G} p(\tau_{ij} ) f(x+\tau_{ij})\,.
\end{align}
Thus, using Fourier analysis on Abelian groups, the eigenvalues of the convolution can be calculated \cite{diaconisGroupRepresentationsProbability1988}. For this, instead of a general function $f$, we will use the one-dimensional representation, or character of $G$, $\chi_u: G \rightarrow \{\pm1\}$ corresponding to a $u \in \mathbb{F}_2^{4n}$ defined by
\begin{align}
    \chi_u(\tau_{ij}) = (-1)^{\tau_{ij} \cdot u}\,,
\end{align}
where the exponent is the dot product of the two elements modulo 2. Then, the eigenvalue corresponding to $u$ is the Fourier transform over the elements of the group \cite{diaconisGroupRepresentationsProbability1988}:
\begin{align}
    \lambda(u) & = \sum_{\tau_{ij} \in G} p(\tau_{ij}) (-1)^{\tau_{ij} \cdot u} \label{eq: fourier eigval} \\
               & = \sum_{\tau_{ij} \in G} p(\tau_{ij})  \prod_{k=1}^{4n} (-1)^{(\tau_{ij})_k u_k}\,.
\end{align}

With the weight of the transition denoted $|\tau_{ij}|$, in the bitflip channel, we can express the transition probability as
\begin{align}
    p(\tau_{ij}) & =p^{|\tau_{ij}|} (1-p)^{4n-|\tau_{ij}|}                          \\
                 & = \prod_{k=1}^{4n} p^{(\tau_{ij})_k} (1-p)^{1-(\tau_{ij})_k} \,.
\end{align}
As the group elements in $G$ cover all $2^{4n}$ bitstrings, the eigenvalue factorizes to bitwise products:
\begin{align}
    \lambda(u) & =\sum_{\tau_{ij} \in G}\prod_{k=1}^{4n} p^{(\tau_{ij})_k} (1-p)^{1-(\tau_{ij})_k}(-1)^{(\tau_{ij})_k u_k} \\
               & =\prod_{k=1}^{4n} \sum_{b \in \{0,1\}} p^{b} (1-p)^{1-b} (-1)^{b u_k}                                     \\
               & =\prod_{k:u_k=0} \sum_{b \in \{0,1\}} p^{b} (1-p)^{1-b} \times \nonumber                                  \\
               & \qquad  \prod_{k:u_k=1} \sum_{b \in \{0,1\}} p^{b} (1-p)^{1-b} (-1)^{b}                                   \\
               & =(1-2p)^{|u|}\,.
\end{align}
The maximum eigenvalue is $\lambda(\boldsymbol{0})=1$ for both bitflip channels. For $0 \leq p \leq 0.5$, the second largest eigenvalues for the two channels are:
\begin{align}
    \lambda_2^{\rm noQEC} & =1-2p           \\
    \lambda_2^{\rm UQEC}  & =1-2p_3 \ \ \ .
\end{align}
This simple result quite importantly determines that in order for the ranking to stay the same, the second largest eigenvalue modulus of the GLQEC must stay between $1-2p$ and $1-2p_3$ for arbitrary system size.

\begin{figure*}
    \centering
    \includegraphics[width=1\textwidth]{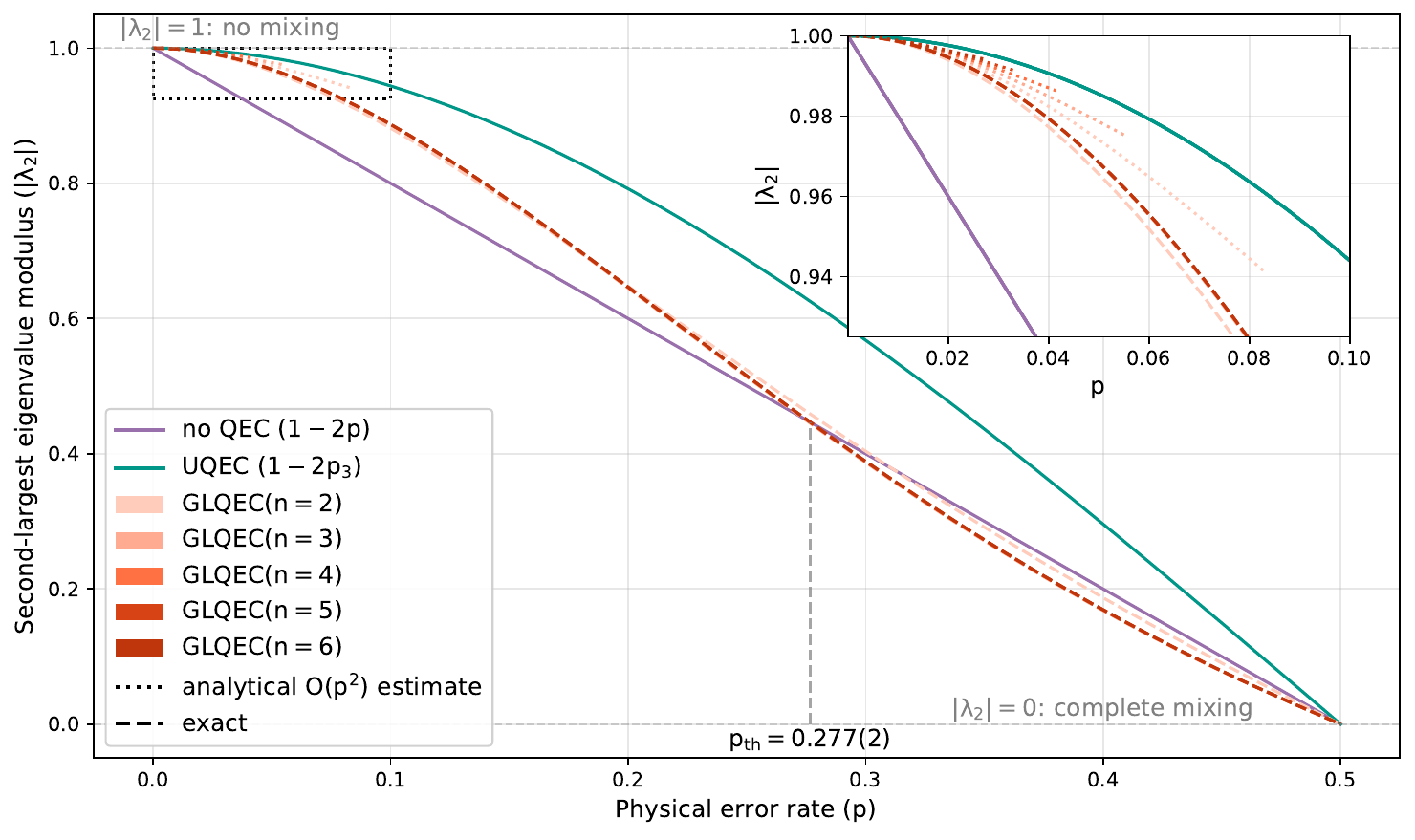}
    \caption[Second largest eigenvalue moduli with PyMatching.]{Exact and estimated $|\lambda_2^{\rm GLQEC}|$ values at different physical error probabilities and $n$ physical sites 2 to 6 using MWPM/PyMatching. Exact values are the eigenvalues of the corresponding classical Markov transition matrix. Estimated (dotted) lines are based on the analytical formulas for $O(p^2)$ truncated Markov chains in \zcref{sssec: analytical lambda2}, drawn up to approximation error $\varepsilon \leq 0.05$. The inset is a zoom of the area marked by the dotted gray rectangle, showing the small $np$ region where the estimates are valid. The exact curves clearly saturate as system size increases for all values of $p$, suggesting a system size-independent mixing speed for the logical GLQEC channel above a certain size. The gray dashed line indicates the threshold $p_{th}$ at which the GLQEC mixing becomes faster than without quantum error correction.}
    \label{fig:exact glqec lambda2 vals}
\end{figure*}

The GLQEC logical channel is similarly classical in nature, with the group of jumps $G^{\rm GLQEC}=\ker H_Z \simeq \mathbb{Z}_2^{2n}$ defined by the parity check matrix $H_Z$ in \zcref{eq: hz parity check matrix}. However, as the GLQEC logical channel contains the recovery operators, the transition probabilities $p^{\rm GLQEC}(\tau_{ij})$ are more complicated than the bitflip channel, and they can depend on the decoding strategy too, not simply on the weight of the jump bitstring $\tau_{ij}$. We can calculate $\lambda_2^{\rm GLQEC}$ exactly by diagonalizing the transition matrix $P^{\rm GLQEC}$ with elements $P^{\rm GLQEC}_{ij}=p^{\rm GLQEC}(\tau_{ij})$ in small cases, up to $n=6$ (24 qubits) as shown in \zcref{fig:exact glqec lambda2 vals}.
We calculate numerically $\lambda_2$ in the interval $0\leq p \leq 0.5$, with 100 points in the $p \in [0, 0.1)$ range and 100 points between $p \in [0.1, 0.5]$. Above $p=0.5$, one can simply invert the bitstring and apply decoding using the inverted bitstring. We observe that $\lambda_2$ rapidly saturates with increasing $n$, suggesting that the mixing speed for GLQEC may exhibit volume independence beyond a certain size, as in bitflip channels. GLQEC mixes faster than UQEC across the whole $0\leq p \leq 0.5$ domain. However, when compared to the noQEC channel, a mixing threshold of $p_{th}=0.277(2)$ is manifest, defined by the interpolated intersection of the $n=6$ GLQEC curve and the noQEC curve, where the uncertainty reported is simply expressing a 0.004 sampling interval.

While \zcref{fig:exact glqec lambda2 vals} looks nearly indistinguishable from its counterpart generated with the extended RRW decoder defined in \zcref{app: extended lookup table decoder}, we observe small deviations in $\lambda_2$ demonstrated in \zcref{tab:glqec_lambda2} of \zcref{app: LUT lambda2} for $p=0.08$. When calculating for a range of $p$ values, this decoder dependence leads to a slightly decreased threshold of $p_{th}=0.261(2)$.

\begin{figure}[htbp!]
    \centering
    \includegraphics[width=1\linewidth]{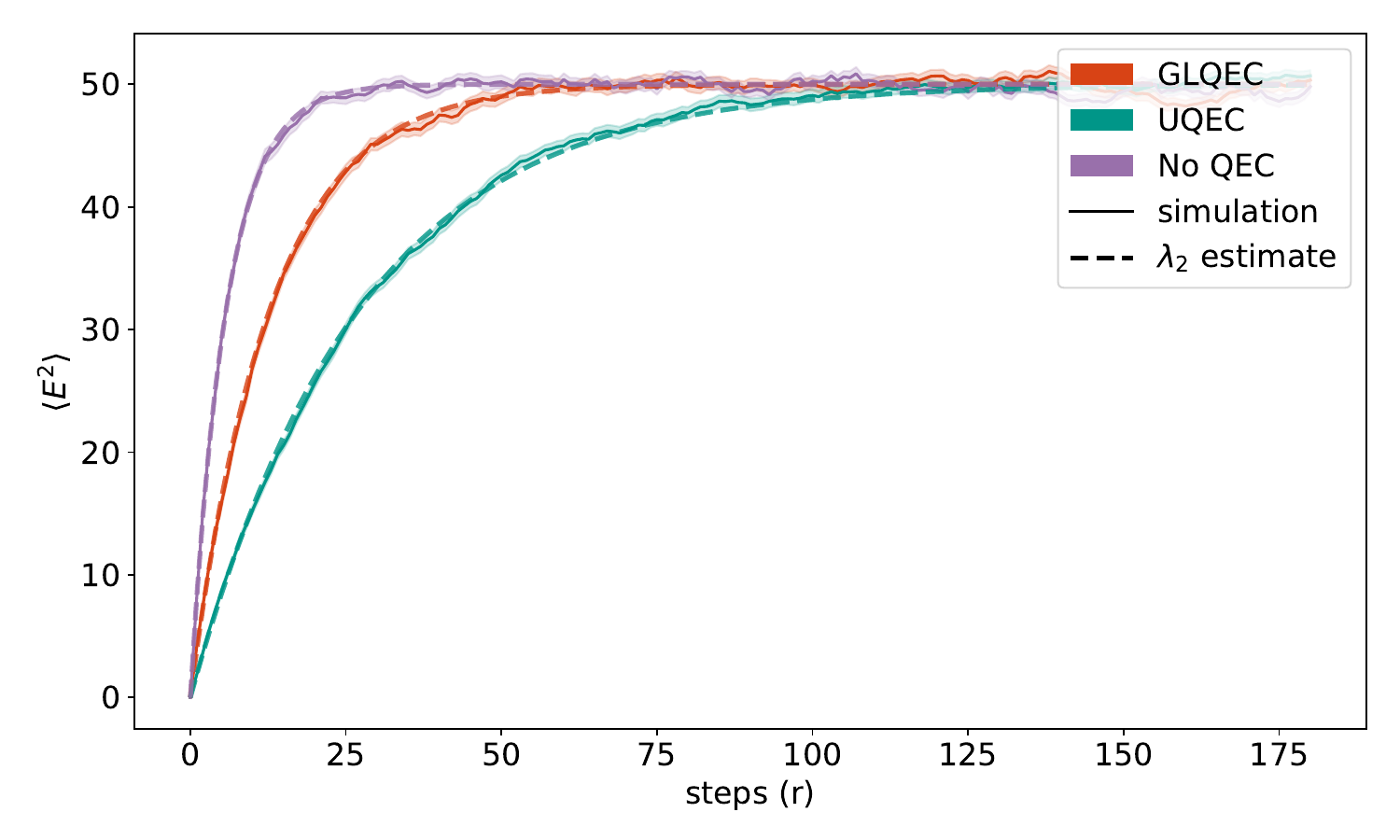}
    \caption[Numerical electric energy estimates for 200 sites.]{Electric energy estimates based on $\lambda_2$ for each channel, showing agreement with Monte Carlo simulation for $n=50$ (200 application qubits). The estimates use $|\lambda_2^{\rm GLQEC}|=\lambda_2^{\rm GLQEC}=0.924416$ based on $n=6$. The good agreement of $\lambda_2$ estimates with the sampled curve confirms the saturation of $\lambda_2^{\rm GLQEC}$ with system size for a fixed physical error rate, in this case $p=0.08$.}
    \label{fig:200slem}
\end{figure}

We test the eigenvalue saturation conjecture by estimating expectation values of $H_E$ using the $\lambda_2^{\rm GLQEC}=0.924416$ value determined at $p=0.08$ for the $n=6$ lattice and compare with Monte Carlo simulated expectation values of a much larger, $n=50$ system. The expectation value defined at time step $r$ as $\bar{\varepsilon}(r)=\Tr[H_E \rho(r)]$, can be modeled as
\begin{align}
    \bar{\varepsilon}(r)\!\sim\! \bar{\varepsilon}(\infty) + \lambda_2^r(\bar{\varepsilon}(0) - \bar{\varepsilon}(\infty))\,,
\end{align}
given the initial $\bar{\varepsilon}(0)=\Tr[H_E \rho_0]$ expectation value and the thermal expectation value $\bar{\varepsilon}(\infty)=\rho_C=\Tr[H_E I_C]$. The GLQEC low-$n$ $\lambda_2$ estimates strongly agree with the high-$n$ Monte Carlo simulated values, demonstrating the saturation as seen in \zcref{fig:200slem}.

Thus, we provided numerical evidence for the ranking of mixing speeds among the three channels. It is interesting to further ask whether we can prove the ranking of the mixing speeds using analytical tools. Next, we make progress in analytically estimating the second largest eigenvalue modulus of the GLQEC channel.

\subsection{Analytical GLQEC SLEM estimates} \label{sssec: analytical lambda2}

The bitflip channel has a simple analytical expression for the second largest eigenvalue as derived in the previous section. For the Gauss's law QEC logical channel, it is practical to construct and diagonalize the $P^{\text{GLQEC}}$ transition matrix to calculate exact eigenvalues only up to a small system size, as displayed in \zcref{fig:exact glqec lambda2 vals}. This process is exponentially costly in system size, because for each transition between logical codewords in $G^{\text{GLQEC}}$, multiple Kraus operators contribute that include a bitflip error of certain weight and the recovery to a logical error as described in \zcref{eq: glqec kraus}. Exactly enumerating all $\tau_{ij} \in G^{\text{GLQEC}}$ requires enumerating all $2^{4n}$ bitstrings, recovering them via a decoder, and grouping them into logical transitions. Here, we show that, for the second largest eigenvalue modulus, we can provide an analytical expression under second-order truncation of the physical bitflip process, only including the dominant contributors to the transitions: zero, one, and two bitflips.

The second-order truncation ignores the probability mass associated with physical bitflips of weight greater than two. Let this approximation error be denoted $m_3$. Then, based on the binomial distribution:
\begin{equation}
    m_3(n,p) = 1 - \sum_{k=0}^2 \binom{4n}{k} p^k (1-p)^{4n-k} \,.
\end{equation}
We can upper bound the missing probability mass (see \zcref{app: falling factorials proof}) by a constant $\varepsilon$
\begin{equation}
    \begin{aligned}
        m_3(n,p) \leq \binom{4n}{3} p^3 \leq \varepsilon \iff 4np \leq (6 \varepsilon)^{1/3} \\
        \text{ for large } n\,.
    \end{aligned}
\end{equation}
For example, by setting $\varepsilon=0.05$, the approximation may be expected to perform well in the regime of $np \lesssim 0.167358$. In \zcref{fig:exact glqec lambda2 vals}, we display the values of leading-order analytical $\lambda_2$ estimates derived in this section only for $n,p$ values within this regime. In this small-$(4np)$ regime, we can focus on the logical transition elements and their probabilities generated by up to two bitflips.

The identity element $\tau_{e}=\boldsymbol{0}$ consists of the bitflip events that are successfully corrected. As the present Gauss's law code is distance 3, all zero-bitflip and single-bitflip events will be included, along with the select two-bitflip events that are successfully correctable. To simplify our discussion, assume $n\geq 3$ from now on, such that the number of correctable order-$p^2$ events is  $C_{n,2}=\binom{4n}{2}-10n$ from \zcref{eq: num correctable weight-2}. In terms of probability masses $\omega_k \equiv p^k(1-p)^{4n-k}$ up to weight $k=2$, the diagonal of the transition matrix, corresponding to the identity element in $G^{\rm GLQEC}$, is $p_0 \equiv P_{i,i}^{\rm GLQEC}$ for all $i$, with
\begin{equation}
    p_0  =  w_0 + 4n w_1 +\left(\binom{4n}{2} - 10 n\right)w_2 \,.
\end{equation}
The non-trivial logical transitions are solely driven by misdecoded second-order events. Altogether, there are $10n$ of these (for $n\geq 3$), but it is important to understand the specific transitions so that we can assess their probabilities. Under minimum-weight decoding, a weight-2 bitflip is decoded using either a weight-1 or a weight-2 recovery operator. When the recovery is incorrect, it results in a weight-3 or weight-4 logical transition. Remember that transitions are elements of the kernel of $H_Z$ (defined in \zcref{eq: hz parity check matrix}). By inspection, we can see that there are $2n$ weight-3 and $2n$ weight-4 elements defined by:
\small
\begin{align}
    \begin{alignedat}{2}
         & \left\{
        \begin{array}{cccccccccc}
            \ldots         & S_{i} & L_{i,i+1} & S_{i+1} & L_{i+1,i+2} & S_{i+2} & \ldots         \\
            \boldsymbol{0} & 1     & 1         & 1       & 0           & 0       & \boldsymbol{0}
        \end{array}
        \right\}_{i=0}^{2n-1}\,,\text{ and} \\
         & \left\{
        \begin{array}{cccccccc}
            \ldots         & S_{i} & L_{i,i+1} & S_{i+1} & L_{i+1,i+2} & S_{i+2} & \ldots         \\
            \boldsymbol{0} & 1     & 1         & 0       & 1           & 1       & \boldsymbol{0}
        \end{array}
        \right\}_{i=0}^{2n-1} \,,
    \end{alignedat}
    \label{eq: w3 and w4 transitions}
\end{align}
\normalsize
respectively. For the weight-3 case, any of the three possible weight-2 bitflip errors will be recovered to the given transition, thus the probability mass of weight-3 transitions is $W_3=3 w_2$ under second-order truncation. Of the $\binom{5}{2}$ weight-2 errors experienced across 5 neighboring application qubits spanning three sites, only 4 have errors not localized to 3 neighbors: (01001), (10010), (10001), and (01010). The first pair and second pair have the same Gauss's law syndrome, allowing one of each to be properly corrected and the other to be recovered to the weight-4 element of \zcref{eq: w3 and w4 transitions}. For example, (01001) and (10001) are the corrected two when using the extended RRW decoder. Thus, the probability of weight-4 transitions induced by weight-2 bitflips is $W_4=2 w_2$. It is instructive that the total probability mass is $2nW_4 + 2nW_3=10nw_2$, which is consistent with the number of incorrectable weight-2 flips calculated earlier (for $n \geq 3$). We are ready to calculate eigenvalues.

We found the identity element $\boldsymbol{0}$, $2n$ weight-3 transition elements ($\in T_3$), and $2n$ weight-4 transition elements ($\in T_4$) in $G^{\rm GLQEC}$, and subsequently calculated their transition probabilities based on only up to weight-2 bitflips. Under this truncation, we set the probabilities for other elements to 0. This results in a substochastic process, for which the eigenvalue calculation can be done via the Fourier transform in \zcref{eq: fourier eigval} for each character defined by an element $u \in \mathbb{F}_2^{4n}$:
\begin{equation}
    \begin{aligned}
        \lambda(u) & = \sum_{\tau_{ij} \in G} p(\tau_{ij})(-1)^{\tau_{ij} \cdot u}                                        \\
                   & =  p_0  +  \sum_{\tau_{ij} \in T_3} 3w_2(-1)^{\tau_{ij} \cdot u} +                                   \\
                   & \qquad \sum_{\tau_{ij} \in T_4} 2w_2(-1)^{\tau_{ij} \cdot u}\,. \label{eq: truncated fourier eigval}
    \end{aligned}
\end{equation}
Now, we can see that for $u=\boldsymbol{0}$, we get the largest eigenvalue, with all $(-1)^{\tau_{ij} \cdot u}=1$, which will equal $\lambda(\boldsymbol{0})=1-m_3(n,p)$. This is the equivalent of the largest eigenvalue being 1.0 for the fully stochastic, non-truncated case. We renormalize with $\lambda(\boldsymbol{0})$, thus $\tilde\lambda(\boldsymbol{0})=\lambda(\boldsymbol{0})/\lambda(\boldsymbol{0})=1.0$.

To calculate the second largest eigenvalue modulus, we first note that all eigenvalues are positive. To see this, a non-negative trivial lower bound (though with no $u$ capable of achieving it) can be calculated by setting all terms besides $p_0$ negative in \zcref{eq: truncated fourier eigval} for $n \geq 3$. Thus, the second largest eigenvalue is equivalent to the second largest eigenvalue modulus.

The highly regular distribution of the weight-3 and weight-4 elements in \zcref{eq: w3 and w4 transitions} allows us to identify the $u$ corresponding to the second largest eigenvalue. This element is $u=(\boldsymbol{0}\ldots0 10\ldots\boldsymbol{0})$, which overlaps exactly with a single weight-3 element at a link $L_{i,i+1}$, and exactly two weight-4 elements. Thus, for these logical transitions, the character coefficient will evaluate to -1. From this, we can analytically calculate the second largest eigenvalue of the truncated process:
\begin{equation}
    \begin{aligned}
        \lambda_2^{\rm GLQEC} & =(1 - p)^{4n} + (4 n) (1 - p)^ {4n - 1} p +                 \\
                              & \quad\left(\binom{4n}{2} - 10 n\right)p^2(1 - p)^{4n - 2} + \\
                              & \quad(2 n - 2) 3 p^2(1 - p)^{4n - 2} +                      \\
                              & \quad(2 n - 4)  2 p^2(1 - p)^{4n - 2}\,.
    \end{aligned}
\end{equation}
Expanding $\tilde\lambda_2^{\rm GLQEC} = \lambda_2^{\rm GLQEC} / \lambda(\boldsymbol{0})$ and the second largest eigenvalue of the noQEC and UQEC channels to second order, we see:
\begin{equation}
    \begin{aligned}
        \tilde\lambda_2^{\rm GLQEC} & = 1 - 14  p^2 + O(p^3) \\
        \lambda_2^{\rm noQEC}       & = 1 - 2p               \\
        \lambda_2^{\rm UQEC}        & = 1 - 6p^2 + O(p^3)\,.
    \end{aligned}
\end{equation}
Thus, for the small-$np$ regime, $\lambda_2^{\rm noQEC} < \tilde\lambda_2^{\rm GLQEC} < \lambda_2^{\rm UQEC}$ holds, which leads to the expected ranking of mixing times for these processes.

We thus found that the dominant contributing processes in the small-$np$ regime demonstrate mixing times that follow the same ranking as we observed numerically. However, the rapid growth of the residual probability mass with increasing $n$ or $p$ severely constrains the applicability of this approximation to the small-$np$ regime.
In particular, this result does not reproduce the saturation of $\lambda_2^{\rm GLQEC}$ demonstrated in \zcref{fig:exact glqec lambda2 vals}. To reproduce the saturation behavior and to explore higher $np$ regimes, analysis of bitflips with weights greater than two is required.

\section{Discussion} \label{sec: discussion}

Leveraging built-in symmetries for quantum error correction in lattice gauge theory simulations is an appealing idea that we numerically and analytically explored in this work. The previously established RRW strategy~\cite{rajput_quantum_2023,spagnoli_fault-tolerant_2024} saves on the number of qubits, making it an attractive target for near-term experiments. However, we found that, compared with an application-agnostic universal QEC code (UQEC) of the same $d=3$ distance, the savings in qubit overhead provided by Gauss's law QEC (GLQEC)  come at the cost of trade-offs that depend non-trivially on the number of lattice sites $n$ and the physical error rate $p$.

For three known options to map the 1+1D Schwinger model to finite-sized hardware, we detailed the (highly) truncated $U(1)$ theory with periodic, non-periodic, or $\mathbb{Z}_2$ (\zcref{app: modular}) electric fields. Our dimensionality analysis clearly shows that in 1+1D with a two-level gauge field, no stabilizer code based on Gauss's law will be compatible with the non-periodic truncated $U(1)$ theory. Our conjecture that this result will hold for larger spatial and gauge dimensions is evidenced in \zcref{app: higher dimensional counting} by numerics in $d=3$ qutrit-truncated gauge field and by the recursive structure of the transfer matrix used for dimension counting. However, a definite proof is a target for future work. It is possible that non-local, more complicated measurement schemes might exist that can enforce the physical subspace of the Hilbert space for the $U(1)_d^-$ theories. One plausible direction would be using codeword stabilized codes (CWS) \cite{cross_codeword_2009}, as one can think of the vacuum state as the stabilized state $\ket{\psi}$, and word operators would be the hopping term operators in the Hamiltonian. Unfortunately, the only known general recovery method for CWS codes \cite{li_clustered_2010} still incurs an exponential number of measurements in system size. Thus, the use of GLQEC in lattice gauge theory simulations will likely force either the use of theories with a periodic electric field or the inclusion of unphysical states in the code space, the impact of which requires further research to quantify.

Focusing on the GLQEC-compatible $U(1)_2^\circ$ theory, we established that in single-round memory experiments GLQEC outperforms UQEC by a factor of at most 1.39 for small lattices, reduced to 1.2 at larger volumes. This advantage goes away as $p$ and $n$ increase, though it is found to be present in experimentally relevant regimes of moderate $p$ and $n$. With a polynomial relationship, the lower the physical error rate, the higher the system size can be raised before the advantage decreases below a fixed level. Transfer matrix formalism and generator function theory \cite{flajoletAnalyticCombinatorics2013} enabled our analytical results to be calculated up to system sizes of $n=50\ 000$, which, in itself, is an unusual luxury when studying QEC systems. Generalizing these tools to open boundary conditions, other linear codes, and quantum codes more broadly warrants further exploration.

With the GLQEC advantage in qubits and single-round memory experiments in mind, we expected that the advantage would manifest in Hamiltonian simulation when using the logical channels of GLQEC and UQEC codes, and especially compared with no quantum error correction. However, based on our density-matrix simulations of the logical channel in small systems, we found that using GLQEC increases decoherence rates and can modify the thermal expectation values of certain observables.
At earlier times, with remaining coherence, this observable-dependent distortion can shift GLQEC expectation values either closer to or further away from the noiseless expectation values compared to UQEC. In this paper, we focused on code-capacity-level QEC protocols and perfect Hamiltonian evolution; the effects of circuit-level noise and Trotterization are left for future study. However, in this idealized setting, we identified GLQEC's effect on thermal expectation values and decoherence rates as a fundamental source of distortion.

When utilizing quantum devices to simulate the quantum fields of nuclear and particle physics as strongly-interacting quantum many-body systems, one observable of promising application is the real-time emergence of local thermalization and its connection to entanglement dynamics~\cite{Ho:2015rga, Kaufman:2016mif, Mueller:2021gxd, deJong:2021wsd, Chen:2024pee, Florio:2024aix, Mueller:2024mmk, Ebner:2025pdm, Florio:2025hoc, Hayata:2026rmv}.
For example, rapid thermalization arises in heavy-ion collisions where the thermalization of quark gluon plasma connects to the onset of hydrodynamics~\cite{Berges:2020fwq}.
Due to the increased mixing rate experienced by the GLQEC strategy, additional care may be required if employing application-specific error correction in simulations focusing on quantum thermodynamic processes.

The logical decoherence rate penalty can be traced to the repeated application of the bitflip channel, which also appears in multi-round memory experiments. We studied this effect in great detail, using Fourier analysis of classical Markov chains and numerical Monte Carlo simulations on classical bitstrings, revealing the statistics of systems up to 200 qubits, where the ranking of mixing speeds still holds true for $p=0.08$. Further numerical evaluation up to system size $n=6$ (24 qubits) indicates that $\lambda_2^{\rm GLQEC}$, the second largest eigenvalue of the GLQEC channel governing the decoherence rate, saturates to a constant value for a given $p$. The saturation result is further evidenced by the ability to predict the evolution of the average total energy in the electric field toward its thermal value at $n=50$ using the $n=6$ calculation of $\lambda_2$.

Interestingly, the GLQEC not only generates mixing rates faster than those of UQEC, but also faster than those of noQEC at high $p$. To quantify this transition, we establish a mixing speed threshold of $p_{th}=0.277(2)$, such that $\lambda_2^{\text{GLQEC}} < \lambda_2^{\text{noQEC}}$ when $p>p_{th}$. However, when it comes to universal error correction, $\lambda_2^{\rm GLQEC} < \lambda_2^{\rm UQEC}$ holds up to the threshold of the bitflip code, $p=0.5$, meaning that the mixing speed penalty must always be considered when implementing GLQEC. For the regime where $np$ is small, we analytically show that the mixing speed ranking among the three channels holds, using an order-2 truncation of the bitflip channel. Extending this perturbative analysis to higher orders by considering three or more bitflips analytically is left for future work.

We see that the GLQEC channel's mixing speed depends on the choice of decoding, with the mixing threshold of $p_{th} = 0.277(2)$ being slightly higher for PyMatching than $p_{th} = 0.261(2)$ for the extended RRW decoder of \zcref{app: extended lookup table decoder}. This suggests a further benefit of using a minimum-weight matching algorithm for decoding over the extended RRW decoding. Understanding the source of this benefit is a topic of future work, as is exploring the effect of a randomized decoder, which would pick uniformly at random among equivalent minimum-weight recovery operators.

To develop further insights into the mixing speed penalty, it would be ideal to deconstruct and visualize the Markov processes in an intuitive way. Unfortunately, beyond being limited to small system sizes, all our attempts---using graphs, matrix plots, or probability mass distribution charts---fell short of clearly and consistently capturing the differences between the channels across $p$ regimes. Instead, our view of the mixing speed is well described by the spectral gap, a global property of the channels. Presumably, as direct visualization of the probability transition graph for a single step is inherently local, differences in spectral gaps are hard to capture reliably with it. Thus, we leave intuitive visualization as an open challenge.

In this paper, we studied the simplest of LGTs, the Abelian $U(1)$ quantum electrodynamics model in 1+1D. Non-Abelian gauge theories with $SU(2)$ and $SU(3)$ symmetries are significantly richer and exhibit a non-Abelian Gauss's law. However, using the loop string hadron (LSH)~\cite{raychowdhuryLoopStringHadron2020,kadamLoopstringhadronFormulationSU32023} formulation of Hamiltonian theories, Gauss's law again becomes Abelian. In Ref.~\cite{mathewProtectingGaugeSymmetries2025b} to protect these gauge symmetries, energy penalty terms are added to the Hamiltonian. It is an inspiring possibility to explore the construction of a QEC code based on the Abelian Gauss's law in the LSH formulation and, if possible, to explore whether similar mixing and dimensionality trade-offs will also be found.

It is interesting to ask what other applications have built-in symmetries that can be used for error correction. As an example, locality preserving spin-fermion mappings for interacting fermionic models on a lattice can have ``built-in'' error-detecting or even error-correcting codes \cite{jiangMajoranaLoopStabilizer2019,setia_bravyi-kitaev_2018,derby_compact_2021,chen_error-correcting_2023}. It has been shown that it is possible to modify these codes to be quantum error transmuting \cite{zhang_quantum_2023} where physical noise translates to some admissible fermionic noise on the logical level. Our analysis motivates research into whether the mixing-speed penalty observed for Gauss's law error correction is also present in other application-specific QEC codes.

\section{Source code availability}

All calculated numerics and source code are available for download as a Zenodo package \cite{pato_2026_18704761}.

\section{Acknowledgments}

The authors thank Andrew Nemec for bringing the OEIS to our attention and Kenneth R. Brown and Iman Marvian for useful discussions. This work was supported by ARO/LPS QCISS program (W911NF-21-1-0005) and the NSF QLCI for Robust Quantum Simulation(OMA-2120757). NK acknowledges funding in part from the NSF STAQ Program(PHY-2325080). We acknowledge the use of Mathematica 14, ChatGPT 5.2 to assist in derivations for \zcref{ssec: mixing speed analysis for lambda2,app: analytical runs of 1s count,app: falling factorials proof}, and LLM-enabled Cursor IDE to accelerate software creation for numerical simulations and figures.

\bibliography{references}

\appendix

\section{Modular \texorpdfstring{$\mathbb{Z}_d$}{Zd} theories} \label{app: modular}

The modular $\mathbb{Z}_d$ theory has gauge group $G=\mathbb{Z}_d$, periodic electric fields, and basis states labeled by flux modulo-$d$ organized around zero, similar to the periodic truncated $U(1)$ theory. With $t=\lfloor \frac{d}{2}\rfloor$, the states will take values $\ket{-t},\ldots,\ket{0},\ldots,\ket{d-t-1}$. A typical choice of link operators involves the conjugate clock/shift operator (generalization of the Pauli matrices for qudits). We keep the $E$ operator to be the simple diagonal operator, but introduce its exponentiated, $Q$ operator as the clock operator:
\begin{align}
    U^{\mathbb{Z}_d}_{s,s+1} & = \sum_{m=0}^{d-1} \bigl(\ketbra{m+1}{m}\bigr)_{s,s+1}                         \\
    E^{\mathbb{Z}_d}_{s,s+1} & = \sum_{m=0}^{d-1} m \bigl(\ketbra{m}{m}\bigr)_{s,s+1}                         \\
    Q^{\mathbb{Z}_d}_{s,s+1} & = \sum_{m=0}^{d-1} e^{i \frac{2\pi}{d} m} \bigl(\ketbra{m}{m}\bigr)_{s,s+1}\,.
\end{align}
Note that due to the periodic structure of $\mathbb{Z}_d$, $U^{\mathbb{Z}_d}\ket{d-1}=\ket{0}$ and $(U^{\mathbb{Z}_d})^\dagger\ket{0}=\ket{d-1}$, $U^{\mathbb{Z}_d}$ is unitary. However, the same-link canonical commutation relationships change to:
\begin{align}
    U^{\mathbb{Z}_d}_{s,s+1}Q^{\mathbb{Z}_d}_{s,s+1}(U^{\mathbb{Z}_d}_{s,s+1})^\dagger = e^{-i\frac{2\pi}{d}}Q^{\mathbb{Z}_d}_{s,s+1} \,.
\end{align}
Gauss's law operator will take the following form:
\begin{align}
    G^{\mathbb{Z}_d}_{s} = \left(E^{\mathbb{Z}_d}_{s-1,s}\right)^\dagger e^{-i\hat{\rho}_{s}} E^{\mathbb{Z}_d}_{s,s+1}\,.
\end{align}
Thus, for a state $\ket{\psi}$ with electric charge $\rho_s$ at site $s$, and flux values $m_{s,s+1}$ between site $s$ and $s+1$:
\begin{align}
    G^{\mathbb{Z}_d}_{s} \ket{\psi} = e^{i\frac{2\pi}{d} (m_{s,s+1}-m_{s-1,s} - \rho_{s})} \ket{\psi},
\end{align}
and $G$ stabilizes the physical subspace:
\begin{align}
    G^{\mathbb{Z}_d}_{s}\mathcal{H}_{phys} = \mathcal{H}_{phys} \,.
\end{align}
It is easy to see that the above Gauss's law operator definition is then equivalently capable of identifying the physical subspace as the $d$-modular version of \zcref{eq: gauss law op in u(1)}.
In order to converge to the electric term of the $U(1)$ Hamiltonian \zcref{eq: electric term} in the $d\rightarrow\infty$ limit, we can choose (following \cite{spagnoli_fault-tolerant_2024} for 1+1D) the $\mathbb{Z}_d$ electric term to be the following:
\begin{align}
    H_E^{\mathbb{Z}_d}=\lambda_E\sum_{s}\sum_{m}f(m) \bigl(\ketbra{m}{m} \bigr)_{s,s+1}\,,
\end{align}
where $\lambda_E f(m)$ converges to $m^2$ in the $d\rightarrow \infty$ limit. Our choice of $f(m)$ will be:
\begin{align}
    f(m) & = e^{i\frac{2\pi}{d} m} + e^{-i\frac{2\pi}{d} m} - 1                                        \\
         & = \cos\left(\frac{2\pi}{d}m\right) - 1                                                      \\
         & = - m^2 \frac{2\pi^2}{d^2} + O\left(\left(\frac{m}{d}\right)^4\right)\,, \label{eq: taylor}
\end{align}
where we used the Taylor expansion for the cosine function in the last line. This leads to $\lambda_E=-\frac{d^2}{2\pi^2}$, and  the final form of the electric term becomes
\begin{align}
    H_E^{\mathbb{Z}_d} & = -\frac{d^2}{2\pi^2} \sum_s \left(Q_{s,s+1} + Q_{s,s+1}^\dagger -I_{s,s+1}\right) \,.
    \label{eq: final modular h_e}
\end{align}
For the $d=2$ qubit case, $\ket{0}$ will represent $\ket{-1}$ flux and $\ket{1}$ the $\ket{0}$ flux, resulting in the following operators:
\begin{align}
    E^{\mathbb{Z}_2}_{s,s+1} & = - \bigl(\ketbra{0}{0}\bigr)_{s,s+1}          \\
    Q^{\mathbb{Z}_2}_{s,s+1} & =  -Z_{s,s+1}                                  \\
    U^{\mathbb{Z}_2}_{s,s+1} & = X_{s,s+1}                                    \\
    H_E^{\mathbb{Z}_2}       & =\frac{2}{\pi^2}\sum_s 2Z_{s,s+1}+I_{s,s+1}\,.
\end{align}
This results in slightly different dynamics than the periodic $U(1)$ case.
Note that the Taylor approximation is only valid when $m/d \ll 1$, that is, only the dynamics with local electric energies low relative to the gauge field truncation $d$ will be accurately captured.

\section{Higher-dimensional gauge fields and qudit stabilizer codes} \label{app: higher dimensional counting}

It is straightforward to generalize the flux configuration space graph and the path counting technique in the main text for larger-dimensional gauge fields. For example, for $d=3$, the fluxes can be one of $-1,0,1$. This then translates to two extra configurations compared to the $d=2$ cases:
\begin{align}
    \mathcal{C}_{e^-}^{(3)} & =\mathcal{C}_{e^-}^{(2)}\cup \{(1,0),(1,1)\}  \\
    \mathcal{C}_{e^+}^{(3)} & =\mathcal{C}_{e^+}^{(2)}\cup \{(0,1),(1,1)\},
\end{align}
where we denoted with $\mathcal{C}^{(d)}$ the configuration space corresponding to $d$ dimensional gauge field.
This is true in general, the $A(d)$ matrix for dimension $d$ will be a $(2d-1) \times (2d-1)$ banded matrix. After simple considerations, one can derive the following iterative technique of constructing $A(d+1)$ from $A(d)$:
\begin{equation}
    \left(
    \begin{array}{c|cc}
                   & 0      & 0      \\
                   & \vdots & \vdots \\
        A(d)       & 1      & 0      \\
                   & 1      & 0      \\[6pt]
        \hline
        0~\cdots~1 & 1      & 1      \\
        0~\cdots~1 & 1      & 1
    \end{array}
    \right),
\end{equation}
where we used the same reverse alphanumeric ordering by $(i_s,o_s)$ of the $e^-$ configurations as in \zcref{eq: config counting matrix for d2}.
Then, using \zcref{eq: dim is sum of eigvals}, we can determine the dimension of the physical subspace. Due to the recursive formula, we conjecture that the qudit physical subspace dimension is also not a perfect power.

For stabilizer qudit codes of qudit-dimension $d$, the code space dimension is $\frac{d^n}{|S|}$, where $|S|$ is the number of elements in the stabilizer group \cite{gheorghiu_standard_2014}. If $d$ is prime, this number is a perfect power of $d$. Thus, the extension of $U(1)^-_d$ for symmetric gauge fields in $d=3$ is expected to continue to be dimensionally incompatible with stabilizer codes. For $n$ sites, the $d=3$ series collides with the OEIS series A198636 \footnote{{https://oeis.org/A198636}}. Based on our testing of up to $n=10000$ sites, none of the numbers are perfect powers. Our conjecture is then that A198636 has no perfect powers. For composite numbers, however, the situation is more nuanced and requires further evidence to prove our conjecture that physical dimensions will continue to be incompatible with stabilizer codes.

\section{The extended RRW decoder} \label{app: extended lookup table decoder}

Here, we extend the RRW lookup table decoder to a minimum-weight 1+1D decoder. Notably, the algorithm is non-local. In the main loop of \zcref{alg: extended lut} (line 11-28), we iterate over the sites from left to right. At each site, we use the RRW table \zcref{tab:RRW decoder} to find the local recovery bit, with the extended case of $(1,1,1)$, where we also choose to recover using the left link qubit $L_{s-1,s}$. For each recovery bitflip, we also flip the corresponding syndrome bits to track progress in the recovery toward the physical subspace. Thus, at the end of the algorithm, the syndrome should be 0. It is easy to see that for the $(0,0,1),(1,0,0),(1,0,1)$ cases, the algorithm will correctly find a recovery based on one of the next sites, thus it can skip adding a flip to the recovery. Finally, no flip is needed for $(0,0,0)$ as Gauss's law is satisfied on all three sites.

The main-loop operations are all local. However, before the left-to-right iteration, we scan to the left to find the start of the first run of ones; this is a non-local operation, in the sense that the running time could be $O(2n)$, where $2n$ is the syndrome size. To avoid an infinite loop during scanning left, we first check for the all-one syndrome case, which is also a non-local operation. The need for these can be understood from the example of a syndrome string $(1,1,0,0,1,1)$  in the case of an $n=3$ physical site lattice. Not scanning left to find the start of the run of ones would result in a recovery of weight three instead of the minimal weight of two.

\begin{algorithm}
    \caption{Extended RRW decoder for 1+1D lattices} \label{alg: extended lut}
    \begin{algorithmic}[1]
        \Function{Decode}{$syn$}
        \If{$syn$ is all ones}
        \State \Return $\{L_{0,1},L_{2,3}, \ldots,L_{2n-2,2n-1} \}$
        \EndIf
        \State $start \gets 0$ \Comment{scan left for start of run}
        \While{$syn[start] =1$}
        \State $start \gets (start-1) \bmod n$
        \EndWhile
        \State $recovery \gets \{\}$
        \State $n \gets \Call{Size}{syn}$
        \For{$i \gets 0$ \textbf{to} $n-1$}
        \State $s \gets (start + i) \bmod n$
        \State $l \gets syn[(s-1) \bmod n]$
        \State $c \gets syn[s]$
        \State $r \gets syn[(s+1) \bmod n]$
        \If{$(l, c, r) = (0, 1, 0)$}
        \State $syn[s] \gets 0$
        \State $recovery \gets recovery \cup \{S_{s}\}$
        \ElsIf{$(l, c, r) = (0, 1, 1)$}
        \State $syn[s] \gets 0$
        \State $syn[(s+1) \bmod n] \gets 0$
        \State $recovery \gets recovery \cup \{L_{s,s+1}\}$
        \ElsIf{$(l, c, r)$ in $[(1, 1, 0),(1, 1, 1)]$}
        \State $syn[s] \gets 0$
        \State $syn[(s-1) \bmod n] \gets 0$
        \State $recovery \gets recovery \cup \{L_{s-1,s}\}$
        \EndIf
        \EndFor \Comment{$syn$ is all zero at this point}

        \State \Return ${recovery}$
        \EndFunction
    \end{algorithmic}
\end{algorithm}

The 1+1D case is particularly simple due to the straightforward dependency between the checks in the single-dimensional traversal of the circle of checks. For 2+1D and 3+1D lattices, the algorithm becomes more complex because more edge cases must be handled. To illustrate this point, Fig.~8 of Ref.~\cite{rajput_quantum_2023} shows the complicated two-round error correction procedure based on lookup tables extended to two spatial dimensions [weight of next sentence is specific to the 2-dimensional context here]. Instead, using MWPM based on the $2n$ weight-5 stabilizer generators would be significantly simpler.

\section{The analytical form for \texorpdfstring{$C_{n,k}$}{Cnk}} \label{app: analytical runs of 1s count}

\begin{figure}
    \centering
    \includegraphics[width=0.8\linewidth]{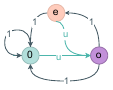}
    \caption{Transfer matrix state machine counting $\sum_j \lceil \frac{\ell_j}{2} \rceil$ in a bitstring with runs of ones of lengths $\ell_j$. The edges should be thought of as multiplicative factors. In the final polynomial for a given bitstring, the power of $u$ counts $\sum_j \lceil \frac{\ell_j}{2} \rceil$. For example, for bitstring $11001101$, the edges are $1 \cdot u \cdot 1 \cdot 1 \cdot u \cdot 1 \cdot 1 \cdot u = u^3$}.
    \label{fig: transfer matrix}
\end{figure}

We want to calculate $C_{n,k}$, the number of runs of ones configurations on $2n$ bits on a cycle that have run lengths $\boldsymbol{\ell}$ and satisfy $\sum_{j} \lceil \frac{\ell_j}{2} \rceil = k$. We will utilize the transfer matrix approach \cite{flajoletAnalyticCombinatorics2013} to create an analytical formula for $C_{n,k}$. This is very similar to the proof we use for counting the number of physical basis states in the truncated $U(1)$ Schwinger model in \zcref{sec: dim of schwinger}, with a bit more complexity, as we will work with a polynomial instead of scalars. For the transfer matrix, we will utilize a state machine that transitions between different states as a function of the bits scanned left to right, as shown in \zcref{fig: transfer matrix}. For a given bitstring, we start from any place and translate to the following states:
\begin{itemize}
    \item the $0$ state: means that we currently read a 0 bit.
    \item the $o$ (odd) state: means that currently we are in a run of 1s and the current number of 1s (on a cycle) counted from the left is odd, or we started anywhere in the all-1 string.
    \item the $e$ (even) state: means that currently we are in a run of 1s and the current number of 1s (on a cycle) counted from the left is even, or we started anywhere in the all-1 string.
\end{itemize}
Note that we will have to account for double-counting the all-1 string later.
We scan the $2n$ bits, and multiply the edges defined by the state machine in \zcref{fig: transfer matrix}. For example, bitstring $11001101$ translates to a state sequence of $eo00oe0o$, which accumulates a factor of $1 \cdot u \cdot 1 \cdot 1 \cdot u \cdot 1 \cdot 1 \cdot u = u^3$.

Given a run of ones of length $\ell_j$, index the position of ones as ${1,\ldots, \ell_j}$, starting from the leftmost one in the run, if it is flanked by zero, otherwise, starting from any of the ones in the run. Then, $\lceil \frac{\ell_j}{2} \rceil$ counts the number of odd positions in the indexing of the run, as both an even $\ell_j=2m$ and $\ell_j=2m-1$ will have $m$ odd positions. This is why we multiply by $u$ only when transitioning into the odd state in the state machine (teal edges in \zcref{fig: transfer matrix}), since only the new odd bits in runs contribute to our statistic.

With this setup, we can represent the adjacency matrix of this weighted directed graph as the matrix
\begin{equation}
    \begin{aligned}
        M(u)=\bordermatrix{
        f/s & 0                   & o & e                   \\
        0   & 1                   & 1 & 1                   \\
        o   & \textcolor{teal}{u} & 0 & \textcolor{teal}{u} \\
        e   & 0                   & 1 & 0
        }\,,
    \end{aligned}
\end{equation}
where columns represent \textbf{s}tarting states of 0, $o$, and $e$ states, and rows represent \textbf{f}inal states of $0,o$, or $e$. Of course, we cannot ever go from $o\rightarrow o$ or $e \rightarrow e$ or $0 \rightarrow e$ while scanning a bitstring, and thus those transitions are~0.

We want to count the number of length $2n$ bitstrings going from $0 \rightarrow 0$, $e \rightarrow e$, and $o \rightarrow o$, because we are on a cycle. Thus, we'll need the sum of the diagonal elements that will count these strings. Let the generating function $W_n(u)$ be the sum
\begin{align}
    W_n(u) = \sum_{k=0}^{n} C_{n,k} u^k \,,
\end{align}
such that $C_{n,k}=[u^k]W_n(u)$ are the coefficients of the $u$ polynomial representation of $W_n$. Then, to account for the double counting of the all-1 string of length $2n$ with $k=n$:
\begin{align}
    W_n(u) = \Tr[M(u)^{2n}] - u^n\,.
\end{align}
We can diagonalize $M(u)$ explicitly and calculate eigenvalues $0, \lambda_{\pm}(u)=\frac{1\pm \sqrt{1+8u}}{2}$, thus
\begin{align}
    W_n(u) = \left(\frac{1-\sqrt{1+8u}}{2}\right)^{2n} + \left(\frac{1+ \sqrt{1+8u}}{2}\right)^{2n} - u^n\,.
\end{align}
We want to extract the coefficients of $u^k$ directly from this expression. Let $s:=\sqrt{1+8u}$. Then
\begin{align}
    (1+s)^{2n} + (1-s)^{2n} & = 2 \sum_{m=0}^n \binom{2n}{2m} s^{2m}\,,
\end{align}
and thus,
\begin{align}
    W_n(u) = \frac{1}{2^{2n-1}}{\sum_{m=0}^n \binom{2n}{2m} (1+8u)^m -u^n}\,.
\end{align}
Now, after another binomial expansion, we find that
\begin{align}
    (1+8u)^m = \sum_{k=0}^m \binom{m}{k}8^ku^k\,.
\end{align}
Finally, using this to extract the coefficient of $u^k$,
\begin{align}
    C_{n,k} & = 2^{3k-2n+1}{\sum_{m=k}^n \binom{2n}{2m} \binom{m}{k}}-\delta_{n,k}
    \,,
\end{align}
where $\delta_{n,k}$ accounts for the $u^n$ term double counting in the case of $n=k$.

\section{Mixing threshold for extended RRW decoding} \label{app: LUT lambda2}

Using the extended RRW decoding defined in \zcref{app: extended lookup table decoder}, we calculate slightly lower $|\lambda_2|$ values for $p=0.08$ compared to using MWPM/PyMatching as shown in \zcref{tab:glqec_lambda2}.

\begin{table}[h]
    \centering
    \begin{tabular}{|c|c|c|c|}
        \hline
        $n$ sites & $N$ qubits & $|\lambda_2|$ (MWPM) & $|\lambda_2|$ (extended RRW) \\
        \hline
        2         & 8          & 0.919475             & 0.918684                     \\
        3         & 12         & 0.924012             & 0.923009                     \\
        4         & 24         & 0.924390             & 0.923270                     \\
        5         & 20         & 0.924414             & 0.923288                     \\
        6         & 24         & 0.924416             & 0.923289                     \\
        \hline
    \end{tabular}
    \caption[Comparing GLQEC second largest eigenvalue moduli for PyMatching and extended RRW decoding]{second largest eigenvalue modulus $|\lambda_2|$ for GLQEC at $p=0.08$ using different decoders. We see rapid convergence in both cases, but slightly lower values for the extended RRW decoder.}
    \label{tab:glqec_lambda2}
\end{table}

\section{Error bound for second-order truncation} \label{app: falling factorials proof}

The number of bitflips in the physical channel of the present lattice gauge theory with $n$ sites can be considered to be a random variable with the binomial distribution $X \sim Bin(4n,p)$, with probability mass function $\Pr(X=k) = \binom{4n}{k} p^k (1-p)^{4n-k}$. Given the truncation of the channel that includes only up to two bitflips, $X \in \{0,1,2\}$, the missing mass is
\begin{align}
    m_3(n,p) = \Pr(X \geq 3)\,.
\end{align}
From $X \geq 3$, it is clear that $X(X-1)(X-2) \geq 6$, which Markov's inequality leads to
\begin{equation}
    \begin{aligned}
        m_3(n,p) & =\Pr(X(X-1)(X-2) \geq 6)                                           \\
                 & \leq \frac{\mathbb{E}[X(X-1)(X-2)]}{6} \,. \label{eq: markov ineq}
    \end{aligned}
\end{equation}
We can rewrite the up-to-three bitflips random variable as a sum over the product of all possible three independent Bernoulli events with probability $p$, thus:
\begin{equation}
    \begin{aligned}
        \mathbb{E}[X(X-1)(X-2)]=\sum_{\substack{i,j,k \\ \text{distinct}}} \mathbb{E}[B_i B_j B_k]\,.
    \end{aligned}
\end{equation}
By the independence of each bitflip, each term $\mathbb{E}[B_i B_j B_k]=p^3$, thus
\begin{align}
    \mathbb{E}[X(X-1)(X-2)] = 3! \binom{4n}{3}p^3 = 6 \binom{4n}{3}p^3 \,.
\end{align}
Referring to \zcref{eq: markov ineq}, the error bound becomes
\begin{equation}
    \begin{aligned}
        m_3(n,p) & \leq \binom{4n}{3} p^3                           \\
                 & =  \frac{4n(4n-1)(4n-2)p^3}{6}                   \\
                 & \approx \frac{(4np)^3}{6}\text{ for large } n\,.
    \end{aligned}
\end{equation}

\end{document}